\documentclass[preprint]{aastex}

\slugcomment{To appear in the Astrophysical Journal,
 February 20, 2001}

\shortauthors{Gott et al.}
\shorttitle{Median statistics}

\newcommand{\omegam}{\Omega_{\rm M}}
\newcommand{\omegal}{\Omega_{\Lambda}}

\newcommand{\kmsmpc}{{\rm \, km\, s}^{-1}{\rm Mpc}^{-1}}

\begin{document}

\title{Median Statistics, $H_0$, and the Accelerating Universe}
\nopagebreak
\author{J. Richard Gott, III\altaffilmark{1},
        Michael S. Vogeley\altaffilmark{2},
        Silviu Podariu\altaffilmark{3},
        and 
        Bharat Ratra\altaffilmark{3}
       }

\altaffiltext{1}{Princeton University Observatory, Princeton, NJ
  08544, jrg@astro.princeton.edu}
\altaffiltext{2}{Department of Physics, Drexel University, Philadelphia, PA 
19104, vogeley@drexel.edu}
\altaffiltext{3}{Department of Physics, Kansas State University, Manhattan, KS
66506, podariu@phys.ksu.edu, ratra@phys.ksu.edu}

\begin{abstract}
We develop median statistics that provide powerful
alternatives to $\chi^2$ likelihood methods and require fewer assumptions about
the data. Application to astronomical data demonstrates that median 
statistics lead to results that are quite similar and almost as 
constraining as $\chi^2$ likelihood methods, but with somewhat more confidence 
since they do not assume Gaussianity of the errors or that their
magnitudes are known. 

Applying median statistics to Huchra's compilation of nearly all
estimates of the Hubble constant, we find a median value $H_0=67
\kmsmpc$.  Median statistics assume only that the measurements are
independent and free of systematic errors.  This estimate
is arguably the best summary of current knowledge
because it uses all available data and, unlike other estimates, makes
no assumption about the distribution of measurement errors.  The 95\%
range of purely statistical errors is $\pm 2\kmsmpc$.  The high degree
of statistical accuracy of this result demonstrates the power of using
only these two assumptions and leads us to analyze the range of
possible systematic errors in the median, which we estimate to be
roughly $\pm 5\kmsmpc$ (95\% limits), dominating over the statistical
errors.

Using a Bayesian median statistics treatment of high-redshift Type Ia
supernovae (SNe Ia) apparent magnitude versus redshift data from Riess
et al., we find the posterior probability that the cosmological
constant $\Lambda > 0$ is 70 or 89\%, depending on the prior
information we include. We find the posterior probability of an open
universe is about 47\% and the probability of a spatially flat
universe is 51 or 38\%. Our results generally support the observers'
conclusions but indicate weaker evidence for $\Lambda > 0$ (less than
2 $\sigma$). Median statistics analysis of the Perlmutter et al.
high-redshift SNe Ia data show that the best-fit flat-$\Lambda$ model
is favored over the best-fit $\Lambda = 0$ open model by odds of
$366:1$; the corresponding Riess et al. odds are $3:1$ (assuming in
each case prior odds of $1:1$).

A scalar field with a potential energy with a ``tail" behaves like a 
time-variable $\Lambda$. Median statistics analyses of the SNe Ia data do not 
rule out such a time-variable $\Lambda$, and may even favor it over a 
time-independent $\Lambda$ and a $\Lambda=0$ open model.

\end{abstract}


\section{Introduction} 
\label{intro}

Statistics that require the fewest assumptions about the data are often the 
most useful. Gott \& Turner (1977, also see Gott 1978) used median 
mass-to-light ratios for groups of galaxies in comparing with N-body 
simulations to estimate $\omegam$. Median mass-to-light ratios were preferable 
to mean mass-to-light ratios because they were less sensitive to the
effects of background contamination and unlucky projection effects.
At the IAU meeting in Tallinn, Estonia, Ya. B. Zeldovich commented on
this choice.  He noted that in Russia some watches were not made very
well, so when three friends meet they compare the times on their
watches --- one says ``it's 1 o'clock'', the second says, ``it's 5
minutes after 1,'' the third says ``it's 5 o'clock''.  Take the
median!  Perhaps no one has ever stated the benefits of the median
over the mean better than Zeldovich\footnote{ 
The mean after removal
of outliers could also prove useful, but that is another
story. We note that while the mean is the quantity that minimizes the sum of
the squares of the deviations of the measurements, the median is the quantity
that minimizes the sum of absolute values of the deviations of the 
measurements.}.  
In this paper we develop median statistics and apply them to
high-redshift SNe Ia apparent magnitude versus redshift data which
recently provided evidence for an accelerating
universe.  We also apply these statistics to estimates of the Hubble
constant and the mass of Pluto.  

The usual hypotheses made when using data in a $\chi^2$ analysis are that
(1) individual data points are statistically independent, (2) there are
no systematic effects,  (3) the errors are Gaussianly distributed, and
(4) one knows the standard deviation of these errors.
These are four extraordinarily potent hypotheses, which lead to
powerful results if the four conditions are indeed true.  We will
show that even the first two conditions alone can lead to powerful
results --- allowing us to drop the third and fourth conditions\footnote{
The $\chi^2$ method may be generalized to take account of correlations, thus 
dropping hypothesis (1), but this requires knowledge of the covariance matrix. 
While the assumption of Gaussianity is not required for parameter estimation 
by simply maximizing the likelihood of $\chi^2$, this assumption is
required for computing the confidence region of the parameters.}

Recent analyses of supernovae distances by Riess et al. (1998, hereafter
R98) and Perlmutter et al. (1999a, hereafter P99) use all four hypotheses. 
These authors combine apparent magnitude versus redshift data for distant
supernovae with data on nearby supernovae in $\chi^2$
analyses to derive likelihood ratios for different cosmological models
(defined by their values of $\omegam$ and $\omegal$, the nonrelativistic matter
and cosmological constant $\Lambda$ energy density contribution to $\Omega$, respectively). Using some
additional Bayesian assumptions, which we discuss below, R98 conclude
that the probability that $\omegal> 0$ is 99.5\% (using MLCS data for all 16
high-redshift SNe Ia including SN 1997ck at redshift $z = 0.97$). P99 
conclude that $\omegal > 0$ with 99.8\% confidence (their fit C).  
These results rely on the important assumption that the errors are normally
distributed, as is apparent in their derivation of confidence limits in
their $\chi^2$ analyses.  This is a somewhat 
troubling assumption since the errors in corrected supernovae luminosities 
are not likely Gaussianly distributed.
While there seems to be a rather strong upper limit on supernovae
luminosities, there seems to be a longer low luminosity tail
(H\"oflich et al. 1996). This does not directly imply that the errors
in the corrected supernovae luminosities are non-Gaussian, but does
indicate that the population of supernovae could include outliers in
the luminosity distribution that are not as well-calibrated by the
training set.
This is related to the possible
concern that, when corrected supernovae luminosities are calculated
using a training set of order the same size (roughly two dozen for the
R98 MLCS method) as the data set to be corrected, one can never be sure
that one is not encountering some supernovae that are odd and
do not fit the training set.

The limits of assuming a normal distribution are illustrated by 
a penguin parable adapted from a discussion by Hill (1992).  
Suppose one measured the weights of a million adult
penguins and found them to have a mean weight of 100 lbs with a
standard deviation of 10 lbs. Further suppose that the observed
data's distribution fits a Gaussian distribution perfectly.  Of the
million penguins measured, suppose that, consistent with a Gaussian
distribution, the heaviest one weighs 147.5 lbs. What is
the probability of encountering, on measuring the next adult penguin,
a penguin weighing more than 200 lbs?  One might be tempted to
say that it was $P = 10^{-23}$, by simply fitting the normal distribution
and calculating the probability of obtaining an upward 10 $\sigma$
fluctuation.  But this would be wrong.  There could be a second
species of penguin, all of whose adults weighed over 200 lbs which
simply had a population a million times smaller, so that one had not
encountered one yet. In this case, the probability of encountering a penguin
weighing over 200 lbs is $10^{-6}$.  Even data that fits a normal
distribution perfectly cannot be used to extend the range beyond that
of the data itself.  The correct answer, suggested by Hill's (1992)
argument, which does not depend on
assumption (3), is that the probability that the 1,000,001st penguin
weighs more than any of the million penguins measured so far is
simply $P = 1/1,000,001$.  This is according to hypothesis (1) that
all the data points are independent. A priori, each of the 1,000,001
penguins must have an equal chance ($1/1,000,001$) of being the heaviest
one.  Thus, the last one must have a probability of $1/1,000,001$ of
weighing over 147.5 lbs.  Beyond that, the data do not say anything.
This may be relevant in the supernova case. Using a training set of a 
little more than two dozen supernovae to correct 16 distant supernovae,
one might encounter a distant supernova that goes beyond the training set, 
in other words, one that is odd. Indeed, supernovae classifications like Ia
pec, as well as Arp's famous catalog of peculiar galaxies, are 
warnings that in astronomy one does encounter peculiar objects as rare
events.  If we fail to recognize them as a separate class, we may
unduly pollute a mean indicator --- another reason for using median
statistics, which make no assumptions about the distribution and which
are less influenced by such outliers. Clearly, given sufficient 
information about a penguin (supernova), we should be able to identify it 
as a different species (supernova class) and thereby avoid skewing the 
results.  Our concern is what happens when the information is not sufficient.

When Gaussian errors are assumed, one of the great benefits is that
the errors decrease as $N^{-1/2}$, where $N$ is the number of measurements.
Thus, with the 16 high-redshift R98 supernovae one can get estimates of 
$\omegam$ and $\omegal$ that are 4 times as accurate as with a single
measurement.  In this paper we show how median statistics takes advantage of a 
similar $N^{-1/2}$ factor to produce accurate results, even while not relying 
on hypotheses (3) and (4) that the errors are Gaussian with known
standard deviation.  Indeed, as we shall show,
hypotheses (1) and (2) are sufficiently powerful by themselves to
produce results that are only slightly less constraining than those 
from $\chi^2$ analyses that also assume hypotheses (3) and (4), but in
which we may have more confidence because two significant and perhaps
questionable assumptions have been dropped.

In Section \ref{medstats} we outline how median statistics can be used with $N$
estimates, which we illustrate with examples from the Cauchy
distribution and another look at the penguin problem. We apply our methods to 
estimates of the Hubble constant in Section \ref{hubble} and to estimates of
the mass of Pluto in Section \ref{pluto}. In Section \ref{supern} we perform a 
simple binomial analysis of the 16 high-redshift R98 SNe Ia measurements.
In Section \ref{bayes} we present a more complete Bayesian analysis of these 
data. Constraints on $\omegam$ and $\omegal$ from the larger P99 data set are 
discussed in Section \ref{perlmutter}. Section \ref{quint} discusses median 
statistics SNe Ia constraints on a time-variable $\Lambda$. In Section 
\ref{conclude} we summarize our conclusions.

\section{Median statistics} 
\label{medstats}

We assume hypotheses (1) and (2), that our measurements of a given
quantity are independent and that there are no systematic effects.
Suppose we were to take a large finite number of measurements. We will
then assume --- call this related hypothesis (2a) --- that the median
value thus obtained as the number $N$ of measurements tends to
infinity will be the true value. We are thus excluding some ``complex"
distributions, e.g., a symmetric double hatbox model with a gap in
the middle. The accuracy of hypothesis (2a) may be limited by
discreteness in the measurements that prevents the data set from
including the true median (see section 3 for an example of this
problem, in which we analyze the Hubble constant data, which are
tabulated as integer values). An extreme example of discreteness would
be the case of a sample of numbers generated by coin flips in which
heads = 1 and tails = 0. If we obtain 49 1's and 51
0's, then the median is 0 but the 95\% confidence limits
must include both 0 and 1.

If we make a large number of measurements and there are no systematic effects 
we might naturally expect half to be above the true value
and half to be below the true value. So we will suppose that after some very 
large number of measurements, as $N$ tends to infinity, there would be a true 
median (TM). Now by hypothesis (1) each individual 
measurement will be statistically independent, thus, each has a 50\% chance 
to be above or below TM. Suppose we make $N$ independent 
measurements $M_i$ where $i = 1,...,N$.  Where is TM likely to be?  The 
probability that exactly $n$ of the $N$ measurements are higher than TM is 
given by the binomial distribution, $P = 2^{-N} N! / [n!(N - n)!]$, because 
there is a 50\% chance that each measurement is higher than TM and they are
independent.  Thus, if we have taken $N$ measurements $M_i$ and these are later 
ranked by value such that $M_j > M_i$ if $j > i$, then the probability that 
the true median TM lies between $M_i$ and $M_{i+1}$ is
\begin{equation}
  P = {2^{-N} N!\over i!(N - i)!} ,
\end{equation}
where we set $M_0 = -\infty$  and $M_{N+1}= +\infty$. For example, if $N = 16$,
our confidence that TM lies between $M_8$ and $M_9$ is 19.6\%.  
Importantly, the distribution of TM is much narrower that the
distribution of the measurements themselves.  For comparison the
probability that the {\bf next} individual measurement we take will lie
between $M_i$ and $M_{i+1}$ is just
\begin{equation}
  P = {1\over N+1} .
\end{equation}
If we set $r = i/N$ we can define $M(r) = M_i$. Then the
distribution width can be defined by the variable $r$. Any
measurement $m$ may be associated with a value $r$ by the inverse function
$r(m)$ such that $M(r) = m$ with suitable interpolation applied.  In the
limit of large $N$ we find that the expectation value of $r$ for the
next measurement $m$ is $\langle r \rangle = 0.5$ and its standard deviation 
is $\langle r^2 - \langle r\rangle ^2\rangle ^{1/2} = 1/(12)^{1/2}$
(since the distribution is uniform in $r$ over the
interval 0 to 1).  On the other hand, in the limit of large $N$ the
expectation value of $r$ for TM is $\langle r\rangle =
0.5$ and its standard deviation is 
$\langle r^2 - \langle r\rangle ^2\rangle ^{1/2} = 1/(4N)^{1/2}$ (in fact, 
in the limit of large $N$ the distribution in $r$ approaches a Gaussian
distribution with the above mean and standard deviation).  Thus, as we
take more measurements we see that the standard deviation in $r$ of the
TM is proportional to $N^{-1/2}$.  If we use median statistics, we
find that our precision in determining TM (as measured
by the percentile $r$ in the distribution of measurements) improves
like $N^{-1/2}$ as $N$ grows larger. Thus, median statistics achieves the 
factor of $N^{1/2}$ improvement with sample size that we expect from
mean Gaussian statistics.

Statistics of samples drawn from a Cauchy distribution illustrate the
robustness of the median for even a pathological parent population.
If $\theta$ is a uniform random variable in the range from $-\pi/2$ to
$+\pi/2$, then the probability distribution function of $x=x_0 + \tan
\theta$ is a Cauchy distribution, $f(x) \propto 1/[1+(x-x_0)^2]$. 
This is a distribution with infinite variance, thus samples from this
parent population are plagued by extreme outliers. However, the median is 
quite well-behaved and the uncertainty in the median, unlike the variance,
is appropriately narrow. As an example, we generate a sample of 101 uniform
random values of $\theta$, then compute the statistics of the set
$\{x_i\}, i=1,...,101$ where $x=5+\tan \theta$. Using standard formulae
we find that the mean of our sample is $\overline{x}=9.58$, the
standard deviation is $\sigma_x=54.8$, and the standard deviation of the mean 
is $\sigma_{\overline{x}} = 5.45$, thus the 95\% confidence limits on the 
mean are $-0.32 < \overline{x} < 19.48$.
For comparison, the median of this sample is $x_{med}=4.818$ and the 95\%
confidence limits on the true median, following eq. (1), are 
$4.41 < x_{TM} < 5.11$. The median is nearly immune to the
``outliers'' in our test sample, which included $x=-35.17$ and $x=552.57$.

Let us apply median statistics to the previously mentioned penguin problem.
Suppose we measure the mass of 1,000,000 penguins and find that
they follow a normal distribution with a mean of 100 lbs and a
standard deviation of 10 lbs. Thus we have 
\begin{equation}
  {\rm mean} = 100\, {\rm lbs}
\end{equation}
and applying the standard formula the standard deviation of the mean is 
\begin{equation}
  \sigma_{\rm mean} = 10\, {\rm lbs}/(999,999)^{1/2} = 0.01\, {\rm lbs}.
\end{equation}
We would hence deduce with 95\% (2 $\sigma$) confidence that the 
true mean for the population of penguins lies between 99.98 and 100.02 lbs.
But this result will be true only if the distribution in penguin
masses beyond the limits seen in the first million penguins is well
behaved, in particular falling off more rapidly than 1/mass.  Suppose
one penguin in a million weighs 100,000,000 lbs.  Since we have
examined only 1 million penguins, there is an appreciable chance
($P = e^{-1} = 0.38$) that we would have missed one of the supermassive
ones.  Yet, these supermassive penguins make the true mean = 200 lbs.
So, even if the already measured data is well behaved, it is easy to
be fooled by extreme cases falling beyond the observed distribution.
One is less likely to be fooled about the median mass.

In the above example we would deduce that, with $N = 1,000,000$, the
expected $r$ value of TM and its standard deviation
would be 0.5 and 0.0005 respectively.  Thus, we would say with 95\% 
(2 $\sigma$) confidence that TM has an $r$ value
between 0.499 and 0.501.  In other words, we expect the true median
weight of penguins to lie between the weight of the 499,000th and the
501,000th most massive of the million measured penguins.  These are
distributed approximately normally so the 499,000th most massive
weighs 99.975 lbs and the 501,000th weighs 100.025 lbs.  Thus, with
95\% confidence we would say that the true median lies between 99.975
lbs and 100.025 lbs.  Note that these limits are only slightly less
constraining than the 95\% confidence limits derived on the mean
earlier.  Furthermore, these limits are not invalidated by the
supermassive one-in-a-million penguins.  Their existence only changes
TM to 100.000025 lbs.  If one's data points are independent 
and there are no systematic effects, the median value is not going to be 
greatly perturbed by data points lying beyond the range of observed values 
--- whereas the mean can always be significantly perturbed.

In short, the 95\% confidence limits on the true median are not much wider 
than those derived for the mean (assuming a Gaussian distribution), and they 
are more secure since the hypothesis of a Gaussian distribution is dropped.

\section{Hubble Constant} 
\label{hubble}

\subsection{Approaches to Hubble Constant Statistics}

The history of attempts to estimate the Hubble constant invites the
application of statistics that are robust with respect to
non-Gaussianity in the error distribution. Until recently, many
published estimates of the Hubble constant, $H_0 = 100 h\kmsmpc$, differed by
several times their quoted uncertainty range. The most famous historical
contradiction was that between estimates published by Sandage, Tammann
and collaborators, typically $h=0.5\pm 0.05$, and
de Vaucouleurs and collaborators, typically $h=0.9\pm 0.1$.
What should one believe when reputable astronomers have published
values that differ by 4 $\sigma$? If a priori we gave equal weight to
the two groups, our technique would ignore the quoted 
errors and allocate a chance $P=25\%$ that $h < 0.5$, $P=50\%$ that 
$0.5 < h < 0.9$, and $P=25\%$ that $h > 0.9 $ --- which is probably 
reasonable if these were the only available data.

Another approach to determining $H_0$ from a collection of
measurements is to filter out, or at least give smaller weight to,
``wrong'' observations and use only the best published estimates.
``Wrong'' in this context means observational values plus their errors
that are unlikely to prove correct given the other data in hand. Press
(1997) develops an elegant Bayesian technique using this approach,
beginning with 13 reputable measurements of $H_0$, and finds a mean of
$h=0.74$ and 95\% confidence (2 $\sigma$ around the mean) range $0.66
< h < 0.82$.

Our approach to analyzing the same Hubble constant data set is like
that suggested by Zeldovich --- use all the data and take the median.
If we apply our median statistics method to the same set of 13 $H_0$
measurements used by Press (1997), we obtain a median of $h=0.73$ and
97.8\% confidence limits of $0.55 < h < 0.81$. These results are nearly
identical to Press's result, without any assumption of Gaussianity or
even looking at the 13 error estimates of the observers.  Note
that our uncertainties are not symmetric and the range
quoted is not exactly 95\% because we do not assume Gaussianity and
therefore we do not interpolate probabilities between the estimates
(one could use these limits as conservative estimates of the 95\%
limits, since the 95\% confidence region lies somewhere between them).

Our treatment assumes that the rank of the next measurement is random,
thus the next measurement is equally likely to land between any of the
previous measurements (or below/above the smallest/largest).  For
those who would like a Bayesian treatment, this assumed distribution
of the next measurement is our prior for the median, which we multiply
by the binomial likelihood of observing $N$ tails/heads to determine the
probability distribution of the true median given the previous data.

\subsection{331 Estimates of the Hubble Constant}

We now apply our median statistics method to 331 published
measurements of the Hubble constant (Huchra 1999) .  After deleting
four entries in the table from 1924 and 1925 that lacked actual
estimates of $H_0$, the June 15, 1999 version of this catalog
contained 331 published estimates, the most recent dated 1999.458.
These have a large range, including Hubble's early high values (near
$h = 5$) and, on the low end, values as small as $h=0.24$ inferred
from measurements of the Sunyaev-Zeldovich effect in clusters (McHardy
et al. 1990).  However the relative likelihood of the true median as
defined by these measurements and using eq. (1) is very narrow, as
indicated by Figure 1.  The published estimates were tabulated as
integer values in $\kmsmpc$ and there are many identical estimates.
Figure 1 shows the relative likelihood of the true median of $H_0$ in
bins centered on these integral values. The median value of the 331
measurements is $h = 0.67$; arguably this is an extremely reasonable
estimate of the Hubble constant. The 95\% statistical confidence
limits are approximately $0.65 < h < 0.69$, obtained by integrating
over the tails of the binomial likelihood distribution. These are
surprisingly narrow limits. This result illustrates that the
assumptions of independence and lack of systematic errors alone are
very powerful.  These purely statistical uncertainties are certainly a
lower bound on the true errors, both because the entries in Huchra's
table are not independent measurements and because of systematic
uncertainties in various methods for measuring $H_0$.  Below we
discuss the possible impact of systematic effects on our median
statistics estimate of the Hubble constant.

The strong effect of including or removing a small number of estimates
illustrates how the mean can be biased upwards or downwards by a few
extreme values, while the median remains insensitive to these
outliers. The mean of the 331 $H_0$ estimates is $h=0.80$ with 95\%
limits $0.76<h<0.84$, inconsistent with our median statistics.
After excluding the 10 estimates published before Sandage's (1958) paper
that discusses Hubble's confusion of HII regions for bright stars, the
median of the remaining 321 estimates is again $0.67$ with 95\% limits
$0.65<h<0.69$. However, the mean of this culled sample is $h=0.68$
with 95\% limits $0.66<h<0.70$, perfectly consistent with median
statistics. The result seems obvious; removing the systematically high
estimates makes sense because we are aware of systematic errors of the
type that Sandage (1958) points out. A strength of median statistics
is robustness when we lack such knowledge.

For comparison with the median and mean, we find that the
mode of the 331 estimates is $h=0.55$. 11 of the 19 estimates with
this value were published by Sandage, Tammann, and collaborators.

The importance of using the median to estimate the true value of $H_0$
from a sample of estimates becomes apparent when we consider the
arbitrariness of the Hubble relation, $v=H_0r$.  A trivial rewriting
of this relation as $r=\tau_H v$ describes identical physics.  Had the
relation first been written in this form, we would all be trying to
measure the Hubble time, $\tau_H=1/H_0$.  However, using the mean to
estimate these parameters would give inconsistent answers because the
mean of a sample of $H_0$ estimates is not the same as the inverse of
the mean of $\tau_H$ estimates, $\bar{H_0} \neq 1/\bar{\tau_H}$.  For
the sample of 331 $H_0$ estimates, the mean of $H_0$ yields $h=0.80$
but the inverse of the mean of $\tau_H$ yields $h=0.66$ (excluding the
ten pre-1958 estimates yields $h=0.68$ and $h=0.65$, respectively).
In contrast, the median yields identical estimates, thus guaranteeing
that the central values of $H_0$ and $\tau_H$ obey the correct
relation $\tau_H = 1/H_0$.

\subsection{Systematic Effects and Uncertainties in the Median of $H_0$}

The small range of the 95\% confidence interval (statistical errors
only) may cause one to immediately and rightly object that these $H_0$
estimates are neither independent nor free of systematic errors.
Regarding the latter objection alone, we might consider the set of
$H_0$ measurements as 331 ``Russian watches.'' As long as the same
systematic effect does not plague an overwhelming number of the
measurements, the median of the measurements should be relatively
robust (certainly more so than the mean). The lack of independence of
these measurements implies that the same systematic bias might affect
at least one group of published estimates but, again, this should not
strongly affect the median unless this group of estimates is a
significant fraction of the 331.  In fact we find that similar
systematics could affect the majority of the estimates and so we must
evaluate this effect.  Of course, the real concern about independence
and systematic errors is their impact on the confidence intervals,
which are remarkably small.  That the 95\% confidence interval of
purely statistical errors ranges
over $\pm 2\kmsmpc$, while astronomers have long argued over
differences of $\pm 10\kmsmpc$ and larger, merely points out that
systematic effects are the likely dominant source of uncertainty.

In the following discussion we attempt to assess the possible impact
of systematic errors on our median statistics analysis.  Exhaustive
analysis of systematic errors in measurement of the Hubble constant is
obviously beyond the scope of this paper.

Many of the methods of $H_0$ estimation have well-known possible
systematics. First, the overwhelming majority of methods are tied to
the LMC distance scale and/or calibration of the Cepheid
period-luminosity relation. Of the 331 estimates, probably not many
more than 52 (those based on the CMB, the Sunyaev-Zeldovich effect,
and gravitational lensing time delays) are certain to be independent
of the LMC and Cepheid observations. Thus, roughly 84\% of the sample
of estimates could share similar systematic uncertainty.  Many recent
measurements, roughly 130 of the 331 ($\sim 39\%$), are specifically
tied to the HST Cepheid distance scale.  So, the data set clearly
violates the above assumption that no group of measurements subject to
the same systematic effect forms a significant fraction of the sample.
Below we address the impact of this lack of independence.

The ``Cepheid-free'' methods each have their own possible systematic
errors. Estimation of $H_0$ using the Sunyaev-Zeldovich effect in
clusters typically assumes that the gas in clusters is smoothly
distributed; clumpiness in the gas would cause the true $H_0$ to be
lower than estimated.  Clusters are more likely to be included in
optical cluster catalogs and targeted for observation of the S-Z
effect if they are prolate along the line-of-sight; such a projection
effect would cause $H_0$ to be larger than estimated (e.g., Sulkanen 1999).
Using gravitational lens time delays to
estimate $H_0$ requires assumption of a model for the mass distribution in
the lens, which can be non-trivial.  Changes in the mass model have
substantial impact on the derived $H_0$ and could
push the estimates up or down, depending on the assumed model (for
discussion regarding 0957+561 see, e.g., Falco, Gorenstein, \& Shapiro
1991, Kochanek 1991).

Uncertainty in the LMC distance seems likely to be the dominant source
of error in the majority of $H_0$ estimates.  The LMC distance modulus
assumed in many recent $H_0$ analyses (e.g., Mould et al. 2000) is $m-M=18.5$,
corresponding to an LMC distance of 50 kpc.  
This distance modulus has quite large uncertainty; some recently
published values span from $m-M=18.1$ (Stanek, Zaritsky, \&
Harris 1998) to 18.7 (Feast, Pont, \& Whitelock 1998). Because a 
shortening of the distance scale by
$\delta(m-M)=0.1$ corresponds to a 4.7\% upward shift in $H_0$, this
could be a very large effect. When Mould et al. use the histogram of
recent LMC distance moduli to model the effect of this uncertainty,
they infer a possible bias of 4.5\% (in the sense that the true value
of $H_0$ would lie above the estimate arrived at when Gaussian
errors were assumed) and a total uncertainty ($1\sigma$) of
12\% in the value of $H_0$ from combining all the Key Project results.

We can assess the uncertainty in the LMC distance modulus and the
resulting uncertainty in the true median of $H_0$ in the same way that
we analyze the $H_0$ data themselves; we apply median statistics to
the distribution of published $m-M$.  Examining Gibson's (2000)
compilation of 38 recent measurements of the LMC distance, we find
that the median of these is $m-M=18.39$ with 95\% confidence limits
$18.3 < (m-M) < 18.52$. This median lies below the nominal $m-M=18.5$
partly due to the number of recent measurements that use red clump
stars, which typically yield $m-M\sim 18.3$ or smaller (this tail of
smaller $m-M$ is also what causes the possible shift of $H_0$ by 4.5\%
in the modeling performed by Mould et al.).  To estimate the range of
systematic uncertainty in the LMC distance modulus due to different
methods we apply median statistics to these data, but give equal
weight to different methods (as we shall do below with $H_0$).
Grouping the 38 LMC estimates into 11 different methods (taking the
median of estimates among each group), the median of the methods (the
median of medians) is $m-M=18.46$ with 95\% limits $18.26 < (m-M) <
18.64$.  This median agrees with the median of values in the histogram
in Figure 1 of Mould et al. (2000). Relative to $H_0$ estimated with
the nominal LMC distance modulus of $m-M=18.5$, the median statistics
of different methods implies that $H_0$ could be shifted upwards by
1.9\% with a 95\% confidence range that spans from 8.0\% below to 9.6\%
above the revised median.

It is reasonable to assume that a range of LMC $m-M$ have been used in
the past; too small a value led to erroneously large $H_0$ and vice
versa. Correcting this ensemble of estimates to use the true value
would therefore narrow the distribution of $H_0$ estimates and might
cause a small shift in the median. However, to evaluate the possible
impact of the LMC distance modulus uncertainty on our median
statistics estimate of $H_0$, let us suppose that all but the 52
``Cepheid-free'' estimates had used the {\it same} value, $m-M=18.5$,
to calibrate the distance scale. Since many workers are know to have
used other distance moduli (e.g., de Vaucouleurs advocated a shorter
distance scale, using $m-M=18.4$ or smaller; de Vaucouleurs 1993),
this assumption may lead us to overestimate the impact on the median.
If all the $H_0$ estimates that could be 
plagued by dependence on the LMC
distance modulus were to shift in identical fashion, then we 
estimate the effect of the 8.0\% lower and 9.6\% upper 
95\% systematic limits by multiplying these bounds by 0.84.
Thus, using the distribution of LMC distance moduli to model the
systematic uncertainty in $H_0$ and assuming that all but 52 of the
estimates suffer from this same systematic uncertainty
yields a possible shift of $\delta h = 0.01$ and 95\% systematic errors of
$(-0.045,+0.055)$, roughly 7.5\% in either direction.

Other systematic effects on the Cepheid distance scale include
metallicity effects on the period-luminosity relation and
uncertainties in photometry.  If metallicity corrections to the
Cepheid zeropoint based on spectroscopic [O/H] abundance (Kennicutt et
al. 1998) are applied to the HST Key Project Cepheids, their summary
estimate of $H_0$ decreases by 4.5\% (Mould et al. 2000). On the other
hand, use of Stetson's (1998) WFPC2 calibration would cause Cepheid
calibration based on HST to be revised in such a way to shift $H_0$
upwards by 4\%. Another photometric uncertainty concerns blending of
Cepheids (Mochejska et al. 2000; cf. Gibson, Maloney, \& Sakai 2000); 
photometric blending of Cepheids with other
stars would cause the distance modulus to be underestimated, thus
overestimating $H_0$. 

Of the four possible sources of systematic error in the HST Cepheid
distance scale that we have mentioned, two (LMC distance modulus,
WPFC2 calibration) might increase $H_0$ while two (metallicity
effects, photometric blending) might decrease $H_0$.  The magnitude of
these effects varies and we do not consider them equally likely, so
assuming mutual cancellation of these effects is not justified.
However, we would be quite unlucky if all these or other systematic
effects fell in the same direction. 
As shown by Mould et al. (2000), the LMC distance scale uncertainty is
the dominant source of systematic error in $H_0$ estimated by the HST
Key Project.
We conclude that the systematic error on the median value of $H_0$ 
(which is not the same as the uncertainty on any one measurement, nor
any one group of measurements, such as those of the HST Key Project)
due to uncertainty in the LMC distance modulus and/or Cepheid calibration is
of order the LMC effect described above, roughly 7.5\% or $\delta h =0.05$
in either direction at the 95\% confidence level. 

Historically, debate among workers in the field has often centered on
the impact of systematic effects on measurement of $H_0$, with the
expected tendency (on which progress in science keenly relies) of each
group to point out systematics that might plague the others'
measurements. We remain agnostic regarding these debates and examine
the possible effects of systematic effects on $H_0$ by analyzing the
distribution of $H_0$ estimates, grouping these estimates by method
and/or by research group. If systematic effects in one method or group
dominate the 331 published estimates, then excluding them should shift
the median statistics estimate.

Huchra classifies the published estimates into 18 primary types by
method and 5 secondary types by author or group of authors.  
Using
these types to group the estimates, we examine the dependence of $H_0$
statistics on the methods employed and the investigators who report
the estimates.  The 5 secondary types, their number, and the median of
estimates in each type are as follows: No Type ($N=216$, median
$h=0.68$), HST Key Project or KP team member ($N=40$, median
$h=0.73$), Theory ($N=3$, median $h=0.47$), Sandage, Tammann, and
collaborators ($N=51$, median $h=0.55$), de Vaucouleurs or van den
Bergh, and collaborators ($N=21$, median $h=0.95$). The median of ``No
Type'' is $h=0.68$, thus excluding results published by the
best-known workers in the field would have no impact on the median
value of $H_0$.
The median of the type medians is also $h=0.68$.
One might also be curious about the effect of
excluding a particular group's work. {\it Excluding} each group in
turn renders the following medians and 95\% confidence limits: no HST KP
$h=0.65$, $0.62<h<0.68$; no Sandage and Tammann $h=0.68$, $0.65<h<0.70$;
no de Vaucouleurs and van den Bergh $h=0.66$, $0.64<h<0.69$. Thus,
completely excluding any one of these renowned investigators or teams
would shift the median by at most $\delta h=0.02$.

The history of a number of fundamental constants 
shows a systematic trend with time. Certainly this is the
case when we compare very high early estimates of $H_0$ with more modern
values, because of systematic effects like those pointed out by
Sandage (1958). Has such a trend continued? Excluding measurements
before 1990 yields a median value $h=0.65$ and 95\% limits $0.62<h<0.68$.
Further culling the sample to include only ``HST era'' measurements
(post 1996) yields a median $h=0.65$ with 95\% limits $0.62<h<0.67$.

We can extend this ``historical'' analysis by examining only
measurements too recent to have been included in the original version
of this manuscript.
Between June 15, 1999 and August 2, 2000, 46 entries were added to
Huchra's catalog, one of these being the value of $h=0.67$ in the
preprint of this paper. How would our analysis treat an entry such as
ours as it appears in Huchra's table? We would take the central value
seriously, but ignore the quite small uncertainty.
Excluding our own value, the median of the remaining 45 new entries is
$h=0.69$ with 95\% confidence limits of $0.65<h<0.71$. The mean of
these same entries is $h=0.67$ with 95\% confidence limits
$0.65<h<0.70$. Thus, comparison with our estimate of the median above shows
that the 45 new estimates are entirely consistent with the
median of the previous 331.

Systematic differences between the results of different methods
of measuring $H_0$ are also apparent in these data.  Grouping the
estimates by method and applying median statistics yields an estimate
of the range of systematic errors that separate the methods.
This approach also addresses the possible concern that
our analysis of the 331 estimates gives
equal weight to each publication, including proceedings and summary
articles that restate previous results.
Table 1 lists the primary types into which
Huchra (1999) classifies the published estimates. Columns 2 and 3 list
the number in each type and the median of estimates for each type
(mean of the central two for even numbers of estimates).

The median of the 18 methods is $h=0.70$, slightly higher than the
median of all 331 estimates.  The 95\% uncertainty range
$0.645<h<0.745$ of the median of methods includes the median of all
331 estimates, $h=0.67$.  This result is unchanged by excluding the
questionable ``Irvine'' (not a method, but rather a meeting), ``No
Type'' and ``CMB fit'' values.  If the median value for each method is
an accurate representation of the result of applying that method, then
these confidence limits on the median of methods is indicative of the
range of systematic error among different methods, roughly 7\% or
$\delta h = 0.05$ in either direction.  It is improbable that the
systematic errors in the various methods all go in the same direction,
therefore correction of systematic errors in all the methods is likely
to narrow the distribution of $H_0$ estimates and might shift the
median.  Some of the systematic spread in the methods may be due to
different assumed LMC distance moduli (so the 7\% systematic range
here is not independent of the uncertainty due to the LMC distance
discussed above) or freedom from that distance scale calibration
(allowing the
``global'' methods such as gravitational lensing and S-Z tend to yield
smaller estimates of $H_0$ than the locally-calibrated methods).  This
range of systematic error is slightly smaller than the 7.5\% range
from possible LMC distance modulus/Cepheid calibration uncertainties
computed above. We expect this to be so, because the 7.5\% range
assumed that all but 52 of the 331 estimates would suffer from
identical systematics, which is clearly an overestimate.

We conclude that the true median of 331 estimates of $H_0$
is $h=0.67\pm 0.02 (95\% \ {\rm statistical}) \pm 0.05 (95\% \ {\rm
systematic})$, where the systematic error, also derived using median
statistics, is dominated by uncertainty in the LMC distance modulus.
This allows for a systematic error range that is slightly
larger than that inferred by examining the range of estimates produced
using different methods of $H_0$ measurement.
Our estimate of
$h=0.67$ is arguably the best current summary of our knowledge of
$H_0$. It is in reasonable agreement with most recent estimates, is
based on almost all measurements, and makes no assumptions about the
distribution of errors from individual measurements.

\section{Mass of Pluto} 
\label{pluto}

The history of mass estimates for Pluto is an extreme example
illustrating the
effects of systematic errors. Early measurements of the mass of the
Pluto-Charon system were obtained by observing perturbations in the
orbit of Neptune. Errors in the orbit of Neptune dominated the
analysis and these were mistaken for the influence of Pluto.  These
errors led to many measurements of Pluto's mass that were larger than
an Earth mass.  This was, of course, a systematic error.  Later, when
Charon was discovered, the mass of the Pluto-Charon system could be
measured with great accuracy.  If we examine 60 published values of
the mass of Pluto-Charon (Marcialis 1997) we obtain a median mass of
approximately 0.7 Earth masses with 95\% confidence limits between 0.1
and 1.0 Earth masses (see Figure 2). This is incorrect because of a
now well-known systematic error, similar to the mistake made by Hubble
in his estimates of $H_0$.  If we examine only the 28 measurements
taken after 1950 (we pick that date simply to divide the century and
the data set roughly in half), we obtain a median value of 0.00246 Earth
masses, which is almost exactly the modernly-accepted value, with 95\%
limits from 0.00236 to 0.08 Earth masses.  Lacking knowledge of the
Neptune systematic, this extreme trend with time would alone provide a
strong clue that systematic errors dominated the uncertainty in
$M_{\rm Pluto}$. Such a trend would not be readily apparent using the
mean. Even after culling the pre-1950 data the mean is still too
high: $M_{\rm Pluto}=0.157$ with standard deviation $0.060$.
This strongly contrasts with
the case of the Hubble constant in the previous section, in which
excluding the pre-1958 data, which
were contaminated by Hubble's systematic error of mistaking HII
regions for stars, does not change the median. 

The lesson here is that median statistics are more robust than the
mean but are not immune to systematic errors.  The point of examining
these Pluto data is to show a case where even median statistics fail;
there is no magic bullet for faulty data sets.

\section{SNe Ia Data and Binomial Constraints on $\omegam$ in the 
$\omegal = 0$ Model} 
\label{supern}

Recent analyses reported by the High-$z$ Supernova Search Team (R98) and 
the Supernova Cosmology Project (P99) place extremely stringent constraints 
on cosmological models, including evidence for a positive cosmological
constant at the many $\sigma$ level.  It is important to examine the
sensitivity of these results to the use of $\chi^2$ analyses and the
assumptions underlying this approach. Both for this purpose and simply
to demonstrate the use of median statistics, here we apply median
statistics to the R98 high-redshift SNe Ia data.  These data are, with
the exception of SN 1997ck at redshift $z = 0.97\,$\footnote{
Including or excluding this SN does not qualitatively alter the conclusions
(R98; Podariu \& Ratra 2000). It is included in our analyses here.}, 
of excellent quality and the size of the high-$z$ part of this 
data set, $N=16$, lends itself to a clear pedagogical discussion of
median statistics. In what follows we use the MLCS data of R98, and set 
$h = 0.652$ (their calibrated value and consistent with our median). In Section \ref{perlmutter} below, we apply our median statistics
analysis to the larger set of high-$z$ SNe Ia from P99.

In the following analyses we use the most recent R98 and P99 data to
constrain cosmological parameters.  We emphasize that, like analyses
done by R98 and P99, our median statistics analyses rely on hypothesis
(2), that there are no systematic effects in the data.  A number of
astrophysical processes and effects (the mechanism responsible for the
supernova, evolution, environmental effects, intergalactic dust, etc.)
could, in principle, strongly affect our conclusions (see, e.g.,
Aguirre 1999; Drell, Loredo, \& Wasserman 2000; Sorokina, Blinnikov, \& 
Bartunov 2000; Wang 2000; H\"oflich et al. 2000; Simonsen \& Hannestad 1999; 
Totani \& Kobayashi 1999; Aldering, Knop, \& Nugent 2000).

We note that our estimate of $H_0$ in section \ref{hubble} is
consistent with that found by R98 from an analysis of their MLCS data,
$h = 0.652 \pm 0.013$ (1 $\sigma$ statistical error only), thus we use
the R98 value of $h=0.652$ in the likelihood analysis of the
supernovae below in order to vary only the statistical method applied
to these data. 

Using each supernova observation to estimate $\omegam$, we use our
median statistic method to obtain a robust estimate of $\omegam$.
Let's first consider the case where we assume that $\omegal = 0$;
these are Friedmann big bang models characterized by the value of
$\omegam$. Each of the 16 distant supernovae produces
an independent estimate of $\omegam$ --- the value of $\omegam$ such
that the supernova's estimated brightness (from looking at the shape
of its light curve) and its predicted brightness (given its $z$)
agree exactly.  Presumably, if we did an enormous number of
such measurements, the true median value of $\omegam$ obtained would
give us our best estimate of the true value of $\omegam$, assuming as
always that there are no systematic effects.  Listed in the left
column of Table 2 are the 16 estimates of $\omegam$ from the 16
supernovae -- ranked in order of their value.  In the middle column is
the confidence (using eq. [1] above) that the true median
($\Omega_{TM}$) lies between the corresponding values just above and
just below it in the column on the left.  

Thus, there is a 0.00153\% chance that the value of $\Omega_{TM}$ 
is greater than 5.96, and a 2.78\% chance that $0.0426 < \Omega_{TM} 
< 0.206$, and so forth. The 99.6\% confidence limits on $\Omega_{TM}$
are $-1.60 < \Omega_{TM} < 0.656$.  The $\omegam = 1$, $\omegal = 0$
model is ruled out at the 99.6\% confidence level. This is a dramatic result.  
R98 likewise rule out this model at a similarly high confidence level but we 
have done so without assuming that the errors are Gaussian.

The chance that $\Omega_{TM} < 0$ (which would be unphysical, and
indicate that a simple Friedmann model with $\omegal = 0$ was
inadequate) is between 89.5\% and 96.2\% so we can not say that the
$\omegal = 0$ Friedmann models with $\omegam > 0$ are ruled out at the 95\%
confidence level.  (It would be correct to say that they are ruled
out at the 89.5\% level however.)  This compares with the more dramatic
R98 statement that there is a 99.5\% probability that $\omegal > 0$ and that 
therefore all $\omegal = 0$ models with 
$\omegam > 0$ are ruled out. These data are not sufficient to cause median 
statistics to rule out the $\omegal = 0$ models (with 95\% confidence).

If we argued that we know independently that $\omegam > 0.0426$ from
nucleosynthesis results and masses in groups and clusters of galaxies,
and from large scale structure, then with this additional constraint
we could argue that the acceptable $\omegal = 0$, $\omegam > 0$ models are
ruled out at the 96.2\% confidence level.  This is just slightly above
the 95\% confidence level.

\section{Bayesian Constraints on $\omegam$ and $\omegal$ from 16 R98
High-$z$ SNe Ia}
\label{bayes}

Observational data favor low density cosmogonies. The simplest low-density
models have either flat spatial hypersurfaces and a constant or time-variable 
cosmological ``constant" $\Lambda$ (see, e.g., Peebles 1984; Peebles \& Ratra 
1988; Sahni \& Starobinsky 2000; Steinhardt 1999; Carroll 2000; Bin\'etruy 
2000), or open spatial hypersurfaces and no $\Lambda$ (see, e.g., Gott 1982, 
1997; Ratra \& Peebles 1994, 1995; Kamionkowski et al. 1994; G\'orski et al. 
1998). In this and the next
section we consider a more general model with a constant $\Lambda$ that has 
open, closed, or flat spatial hypersurfaces. Two of the currently favored 
models lie along the lines $\omegal = 0$ or $\omegam + \omegal = 1$ in the 
two-dimensional ($\omegal$, $\omegam$) parameter plane of this more general 
model. In Section 8 we consider a model with a time-variable $\Lambda$. 

We can translate the binomial results (such as those derived in the
previous section) into Bayesian constraints. Bayesian statistics says
that the posterior probability of a particular model after analyzing
the data at hand is proportional to the prior probability of that
model multiplied by the likelihood of obtaining the observational data
given that model.

Consider a model with $\omegal = 0$ and $\omegam = 6$. For this model, all 
16 supernovae estimates of $\omegam$ are lower than the true value 
$\omegam = 6$ (see Table 2), thus all 16 distant supernovae would have
intrinsic luminosities that are fainter than we expect.  Since each
represents independent data and we are assuming no systematic effects,
that means that the likelihood of this happening is $1/2^{16}$ (since each
individual supernova has an independent probability of 1/2 of being
fainter than we expect based on the low redshift supernovae).

Suppose we next consider a model with $\omegal = 0$
and $\omegam = 2$.  Table 2 shows that for this
model there is 1 supernova that is too bright and 15 that are too
faint.  The likelihood of obtaining this result is $16/2^{16}$ according to
the binomial distribution.  The relative likelihood in column 3 of the table
is given as 16.  The normalized likelihood is equal to the relative
likelihood in the table divided by $2^{16}$.  From Table 2 we
would conclude that if we initially found a model with $\omegal = 0$,
$\omegam = 6$ and a model with $\omegal = 0$, $\omegam = 2$ to be a
priori equally likely, with odds of $1:1$, then after consulting the
supernovae data we would give odds of $16:1$ in favor of the $\omegam =
2$ model over the $\omegam = 6$ model.

One can perform similar analyses for models with other values of
$\omegam$ and $\omegal$ by examining Figure 3. These plots show, for
each supernova, the locus of values of ($\omegam$, $\omegal$) that
predict the corrected apparent brightness (see, e.g., Goobar \&
Perlmutter 1995).
To compute the likelihood of a particular model (value of $\omegam$ 
and $\omegal$), count the number of SNe Ia that are too bright/faint for 
the model, compute the binomial likelihoods, and apply the prior (note that 
2 SNe Ia lie off the bottom and 4 off the top of the linear scale plot, Figure 
3$b$). Figure 4 allows one to do this ``by eye''. In this figure, the 
greyscale intensity at each point in the ($\omegam$, $\omegal$) plane is
proportional to the binomial likelihood of the observed SNe Ia being
brighter/fainter than predicted by the model with that pair of values
of ($\omegam$, $\omegal$).  Not surprisingly, the favored region in
this plane is similar to that found by R98. Solid lines in Figure 4
show 1, 2, and 3 $\sigma$ likelihood contours derived from a $\chi^2$
analysis (Podariu \& Ratra 2000).

If we limit ourselves to consideration of flat cosmologies then $\omegam +
\omegal =1$ and the allowed models lie along the long-dashed ``flat universe''
lines in Figures 3 and 4. To examine this region in detail, Figure 5 plots 
the relative likelihoods of $\Omega_{TM}$ lying in ranges of $\omegam$ 
bounded by the intersection of the loci in Figure 3$a$ with the flat universe
line. Irrespective of the assumed prior, the best-fit flat-$\Lambda$ model has 
$\omegam \sim 0.3$, in agreement with R98 and P99.

One must adopt reasonable prior probabilities to perform a more complete 
Bayesian analysis. If the prior probability was $P = 100\%$ that the
$\omegam = 1$, $\omegal = 0$ fiducial Friedmann model was correct,
then no matter what data was examined, after examining that data one
would still conclude with $100\%$ certainty that the $\omegam = 1$,
$\omegal = 0$ model was correct. This is because the prior probability
of all other models is zero, and zero times even a high likelihood is
still zero.  Thus this is a bad prior. Priors should be as agnostic as
possible to allow, as much as possible, the data and not the prior
determine the odds.

Vague or ``non-informative'' priors are appropriate in this situation 
(Press 1989). An appropriate vague prior for an unbounded
variable $x$ that can be positive, zero, or negative is uniform in $x$:
$P(x)dx \propto dx$.  But for an unbounded variable $x$ that must
be positive the correct vague prior is the Jeffreys (1961) prior which is
uniform in the logarithm of $x$:  $P(x)dx \propto d\ln x = dx/x$ (Berger 1985).
(If $x$ must be positive then the variable $\ln x$ can be positive, zero, or 
negative and therefore should be distributed uniformly in $d\ln x$ via the
previous rule.)  That the vague prior for a number that is positive
and unbounded should be uniform in the logarithm is well established
and related to the rule that the first digits of positive numbers in a 
data table (like lengths of rivers) should be distributed according to 
the space they occupy on the slide rule (i.e., uniform in the logarithm). 
Thus in any data set involving positive numbers we should expect as many 
to have a first digit of 1 as the sum of those starting with 2 or 3, and 
as the sum of those starting with 4, 5, 6, or 7.

R98 and P99 assume a vague prior with $\omegam$ and $\omegal$ as free
parameters and the prior probability proportional to $d\omegam
d\omegal$.  This would be appropriate for variables that could take on
positive or zero or negative values.  This may be reasonable for
$\omegal$ since people certainly consider both positive and zero
values of $\omegal$ plausible. Since we know little about what sets
the level of the vacuum density it could conceivably be negative as well. 
However, $\omegam$ must be positive.  No one envisions a negative or
zero density of matter.  In fact, R98 and P99 do not consider models
with $\omegam$ not positive.  Standard Friedmann models with negative
values of $\omegam$ would easily explain the data with maximum
likelihood (e.g., $-0.303 < \omegam < -0.266$ from Table 2) but these are 
not considered because $\omegam \le 0$ is thought to be unphysical.
It is also clear that $\omegam$ is not a priori bounded above.  
Thus, the appropriate vague prior for $\omegam$ is uniform in $\ln\omegam$.

In other words, a priori there should be an equal probability of
finding $\omegam$ between 0.1 and 0.2 or finding $\omegam$ between
0.2 and 0.4. More generally, the log prior allows an equal chance of the 
universe being either open or closed.
So if both $\omegam$ and $\omegal$ are free parameters,
we should expect a priori, before examining the data, for 
$P(\omegal,\omegam) d\omegal d\omegam$ to be proportional to $d\omegal d\omegam/\omegam$.  This is more favorable to low density
models than the prior R98 and P99 have chosen\footnote{
Podariu \& Ratra (2000) illustrate the effect of such a non-informative
prior on the confidence contours derived from $\chi^2$ analyses.}.

Perhaps a more serious problem with the prior adopted by R98 and P99 is
that it gives zero weight to the flat-$\Lambda$ model and the $\omegal = 0$
model. After assuming $P(\omegal, \omegam) d\omegal d\omegam$ proportional to 
$d\omegal d\omegam$, R98 state that the posterior probability that 
$\omegal > 0$ is 99.5\%. But what this 
really means is that, according to their prior, the posterior probability that 
$\omegal < 0$ is 0.5\% and that the probability that $\omegal = 0$ is 0\%. 
Furthermore, the posterior probability that $\omegal + \omegam =1$ is also 0\%. 
This is because the prior probability of $\omegal = 0$ or $\omegal + \omegam = 
1$ is zero because these are lines of zero area in the $\omegal, \omegam$
plane. Clearly, the prior adopted by R98 and P99 is not reasonable. This 
shortcoming of these analyses has also been noted by Drell et al. (2000).

Occam's razor suggests that models that are simpler must have higher
prior probability.  One suggestion often used is that the prior
probability for a model with $N$ free parameters is $P = (1/2)^{N+1}$. 
Thus the prior probability that the correct model
is one with no free fitting parameters is 50\%.  The prior
probability that the correct model is one with one free fitting
parameter is 25\%, and with two free fitting parameters is 12.5\% and so
forth.  The infinite sum, up to $N = \infty$ equals 100\% as
expected.  Having additional free parameters to fit the data always
makes fitting any data easier and there has to be a penalty for this. The 
$\omegam = 1$, $\omegal = 0$ Einstein-de Sitter model is one with no free
fitting parameters.  For this reason it has been called the fiducial
cold dark matter model.  The steady-state model also is a model with no
free fitting parameters --- this was one of the attractions of this model
for its proponents.  The steady-state model is spatially flat and expands
exponentially, $a(t)\propto \exp(t/r_0)$.  Geometrically, this model is 
identical to a $\omegal = 1, \omegam = 0$ model. If the steady-state
model were correct 12 supernovae would be too bright and 4 would be
too faint, giving a likelihood of $1820/2^{16}$.
The $\omegam = 1$ Einstein-de Sitter model by contrast has 14 supernovae too 
faint and 2 supernovae too bright, giving a likelihood of $120/2^{16}$. If 
we regarded the odds between these two competing models as a priori $1:1$ 
before examining the supernovae data, we would give posterior odds of $15:1$ 
in favor of the steady-state model after examining the supernovae data.

Let us illustrate our Bayesian technique by considering how we would
have evaluated the competing models in the early 1960's, if the
supernovae data had been available then.  At that time $\omegal > 0$ models
were not popular.  The two main zero free parameter models were the
$\omegam = 1$, $\omegal = 0$ Einstein-de Sitter model and the steady-state 
model.  The only popular one-parameter model was the $\omegal = 0$ Friedmann 
model with $\omegam$ as a free parameter. This may be considered the Friedmann
models with $\omegam \neq 1$ because in this one-parameter family the
$\omegam = 1$ model is a set of measure zero.  Now, if zero-parameter
models as a group were considered to have prior probability of
50\%, and one-parameter models as a group had a prior probability of 25\%,
then we would assign a prior probability of 25\% to the steady-state 
model, 25\% to the Einstein-de Sitter model, and 25\% to the $\omegal = 0$, 
$\omegam \neq 1$ Friedmann model, and 25\% to more complicated models
with 2 or more free parameters.  If we were to discount more
complicated models and renormalize, then we would have prior
probabilities of 33.3\% for the steady-state model, 33.3\% for the
Einstein-de Sitter model, and 33.3\% for the $\omegal = 0$, $\omegam \neq 1$
Friedmann model.  Independent measurements of the mass in clusters of galaxies
would suggest a minimum value of $\omegam$ of 0.05, and Hubble
diagrams to measure $q_0$ from galaxies indicate a maximum value of $\omegam =
4$.  Since the prior for $\omegam$ is distributed uniformly in
$\ln\omegam$ we can calculate the prior probabilities of finding
$\omegam$ in different ranges.  These prior values will be revised by
the likelihoods after examining the supernova data.  Table 3 lists
the prior probabilities for the different models, and how these values
would be revised by multiplying the priors in each model (and over
each range of $\omegam$ in the $\omegal = 0$ Friedmann models with 
$\omegam \neq 1$) by the relative likelihoods from Table 2 above, and 
renormalizing the results to give a total probability of 100\%.

The steady-state model and the $0.05 < \omegam < 0.2$ models would be the
only ones to gain ground due to the supernovae data.  Ranking the 5
models in order of posterior probability, we would see that at the 95\%
confidence level (that our reduced list still included the correct
model) we could only rule out the $1 < \omegam < 4$ models.  The others would
remain in contention.  The steady-state model would have been favored
by the supernova data.  It is of course an accelerating model,  but
one that is no longer in contention.

Today the models in contention are different.  The steady-state model
has no Big Bang and
is ruled out by the cosmic microwave background.  The only zero-parameter
model still in contention is the Einstein-de Sitter model so by Occam's Razor 
it gets 50\% of the prior probability. There are two one-parameter models in 
contention, the $\omegal = 0$ open model with $0.05 < \omegam < 4$, and 
the $\omegam + \omegal = 1$ flat-$\Lambda$ model with $-1 < \omegal < 0.95$.
Together, these one-parameter models must get 25\% of the prior probability.
The two-parameter model has both $\omegam$ and $\omegal$ variable,
with $0.05 < \omegam < 4$, and $-1 < \omegal < 1$.  $\omegal$ can be
both positive or negative and so its prior probability is uniform in
$d\omegal$. $\omegam$ must be positive and so is distributed uniformly
in $d\ln \omegam$.  This is the only two-parameter model under
consideration so its prior probability must be 12.5\%.  The prior
probability of other more complicated models would be 12.5\%.  We can
renormalize to give unit probability to the sum of just the models
under consideration (with 0, 1, and 2 free parameters).  Summing the
prior probabilities listed in Table 4, we find prior probabilities of 
18.6\% for open, 71.4\% for flat, and 9.96\% for closed.  
We have prior probabilities of 14.5\% for $\Lambda < 0$, 
71.4\% for $\Lambda = 0$, and 14.1\% for $\Lambda > 0$. 

After observing the supernovae, the zero-parameter Einstein-de Sitter model 
suffers greatly, dropping to 9.37\%, though still not ruled out by the usual 
95\% criterion. The greatest beneficiaries of the supernova results are 
$\Lambda>0$ models; flat $\Lambda>0$ models rise from 6.97\% to 41.53\% while
open $\Lambda>0$ models rise from a mere 3.34\% to 27.48\%.
Almost as impressive are quite low $\omegam$ open ($\omegal = 0$)
models, that rise in probability from 4.52\% to 13.33\%.

Table 5 summarizes this analysis. First, let us examine the evidence for a 
non-zero cosmological constant. We find that the posterior probability of 
$\Lambda>0$ is 70\%. This result differs from the 99.5\% claimed by R98.
A posteriori, $\Lambda=0$ models have a 27\% probability of being correct, 
thus such models are still quite viable, in agreement with the conclusion
of Drell et al. (2000). R98 find 0\% probability for such models, because 
they disallowed this possibility in their prior. Similar to R98, we find that 
$\Lambda<0$ models are ruled out at greater than 97\% confidence.

Is the universe open or closed? We rule out closed-universe models with
greater than 98\% confidence, but the odds are evenly split between flat and 
open models. The 16 SNe Ia  slightly decrease the probability of flat models, 
from our prior of 71\% to a posterior probability of 51.5\%, while 
significantly increasing the probability of open models, from our prior of 
18.6\% to a posterior probability of 47\%.

Alternatively, we could be more conservative since, because of age
considerations and the amount of power on large scales in galaxy
clustering, it could be argued that the only models currently under serious
discussion are those with $0.05 \leq \omegam < 1$, and with $0 \leq
\omegal < 1$.  That eliminates all parameter-free models, leaving the
flat-$\Lambda$ ($\omegam + \omegal = 1$) model and the $\omegal = 0$
open model as the only one-parameter ones and allows the two-parameter
model where $\omegam$ and $\omegal$ are both allowed to vary.  The
one-parameter models together have a prior probability that is twice
that of the two-parameter model, again by Occam's razor. Since there
are two competing one-parameter models, all three models must have
equal prior probability of 33.3\%.  This reflects fairly well the
prior probabilities as thought of by astronomers today, before seeing
the supernova data. Again since $\Lambda$ may be zero, we use a prior
that is uniform in $d\omegal$ for both the flat-$\Lambda$ model and
the two-parameter model.  Figure 4$b$ shows the relative likelihood
for models in this more restricted ($\omegam$, $\omegal$) space, with
greyscale intensity proportional to the likelihood as in Figure 4$a$.
For comparison, the plotted lines show the confidence regions computed
in similar fashion to R98 (derived in Podariu \& Ratra 2000).

Table 6 presents the priors and results after including the SNe Ia data
for this restricted modern analysis. This gives prior probabilities of 
56.1\% for open, 33.3\% for flat, and 10.5\% closed.
It also gives prior probabilities of 33.3\% for $\Lambda = 0$, and
66.7\% for $\Lambda > 0$. Under these restricted conditions
the $\omegal = 0$ Friedmann models with $0.2 < \omegam < 1$ 
suffer the most, but end with the same $\sim 3\%$ posterior probability 
as under the broader analysis above. 
Open universe $\Lambda>0$ models substantially increase in probability after
considering the supernova data. This analysis uses more (non-SNe Ia) 
astrophysical evidence in the computation of the prior probabilities, thus 
restricting $\omegam$ and $\omegal$ to smaller ranges than those considered
reasonable in the previous analysis.

So far, we have used a log prior for $\omegam$, in keeping with the
positive-definite property of the density of matter.  To examine the
sensitivity of our median statistics results to this choice of prior
for $\omegam$, we also compute posterior probabilities using a uniform
prior for $\omegam$. This is useful because our analysis above differs
from those of R98 and P99 both in its use of median statistics and
in the choice of prior.  Table 7 repeats the analysis summarized in
Table 4, this time with prior probabilities that are uniform in
$d\omegam\, d\omegal$.  When so little prior probability is assigned
to low values of $\omegam$, low density Friedmann models do not fare as well.
Flat-$\Lambda$ models with $\omegal>0$ fare better (59.2\% vs. 41.5\%) with a 
uniform prior, not because of a larger prior probability nor because of a 
higher average likelihood (both actually decrease somewhat relative to the
log prior case), but rather because the average likelihood of other
models decreases when more weight is given to the high $\omegam$
region of the ($\omegam$, $\omegal$) plane.  The total posterior
probability of all $\Lambda>0$ models is only marginally higher
(76.9\% vs. 70.2\%) than in the log prior case.  Thus, our conclusions about the
cosmological constant are relatively insensitive to the choice of prior for 
$\omegam$. However, the posterior probabilities for the flat and closed 
models are now significantly larger, 75.1\% and 11.1\% respectively
(compared to 51.5\% and 1.5\% in the previous analysis), while the odds 
for the open case are significantly reduced to 13.7\% (from 47\%). If, as in 
R98, we were to adopt a uniform prior and limit ourselves to two-parameter 
models, then after renormalizing the posterior probabilities in Table 7 we 
would find a 94.3\% chance that $\Lambda >0$, comparable to 99.5\% in R98. 
These results are qualitatively consistent with those found from the $\chi^2$ 
analyses of Podariu \& Ratra (2000),
showing that median statistics lead to quite similar (but slightly
more conservative) results while relying on fewer hypotheses.
Again we would argue that our choice of priors is superior to those
chosen in R98 and we have included these last estimates only to show
the direct action of the median statistics.
That some of these results depend 
significantly on the choice of prior indicates that better data are needed to 
convincingly constrain cosmological parameters.

\section{Binomial Constraints on $\omegam$ and $\omegal$ from 42 P99 High-$z$ 
SNe Ia}
\label{perlmutter}

One nice thing about median statistics is that they are extraordinarily 
easy to apply. P99 have recently published data on 42 high-$z$ 
SNe Ia\footnote{
While their primary analysis (fit C) makes use of only 38 of these
SNe, we use all 42 in our analyses here. As discussed in P99,
including or excluding the 4 ``suspicious" SNe does not dramatically
alter the conclusion.}. 
They have shown plots of the data versus several cosmological models, thus 
one can read the answer right off their graph of magnitude residuals versus 
cosmological models (their Figure 2). We ignore the error bars and simply 
ask how many data points are below or above each cosmological model line. In 
other words, we examine how many supernovae are too bright or too faint given 
a particular cosmological model. The results are given in Table 8.
For example, for the $\omegam=1, \omegal=0$ model, 4 supernovae are
too bright and 38 are too faint. If this model is correct, the
likelihood of obtaining this result is the same as throwing up 42
coins and having 4 come up heads and 38 come up tails. The open model, 
with $\omegam=0$ and $\omegal=0$, has 10 supernovae too bright and 32
too faint. This is presumably the best fitting $\omegal = 0$ model. The best 
fitting flat-$\Lambda$ model is, according to P99, one with $\omegam=0.28$ and 
$\omegal=0.72$, where 21 supernovae are too bright and 21 too faint. 
Interestingly, this is also a best fitting model using the median statistic! 
No result is more likely than 21 supernovae too bright and 21 too faint.
The ``steady-state'' model with $\omegam=0$ and $\omegal=1$ has 31 supernovae
too bright and 11 too faint. 

If the number of supernovae too bright is $B$ and the number too faint is 
$F$, then according to eq. (1) the relative likelihood of obtaining this 
result in a given model is proportional to $2^{-42}(42!)/(B!F!)$. Table 8
gives the relative likelihoods normalized to the open model. According
to Bayesian statistics our posterior probabilities for each model
after examining the P99 data would be proportional to our
prior probabilities times the likelihoods in Table 8. Today our prior
probability for the ``steady-state'' model is near zero since it has
no Big Bang and cannot explain the cosmic microwave background. If we
restrict attention to open and flat-$\Lambda$ models we
see that even if the $\omegam=1, \omegal=0$ model is favored a priori
by a factor of 2 because it is a simpler zero-parameter model, it is
still strongly ruled out after examining the P99 data
because the likelihood for this model in Table 8 is so low.

If a priori we regarded the best-fitting flat-$\Lambda$ model and the 
best-fitting open model as equally likely (prior odds of $1:1$), then after 
examining the P99 data we should favor the flat-$\Lambda$ model by odds of 
$366:1$. This is an impressive result and does not assume that the errors
are Gaussian or that their magnitude is known. It does rely on the
assumption that the data points are independent and very importantly
that there are no systematic effects. A modest systematic effect in the 
high-redshift supernovae would reverse these odds. The middle panel of 
Figure 2 in P99 shows that ten SNe Ia lie between the curves for
the $\omegam=0.28, \omegal=0.72$ and $\omegam=0, \omegal=0$ models.
The largest magnitude residual between these curves is approximately
$\Delta m = 0.14$, thus a systematic shift of $-0.14$ mag would cause
the data to strongly favor the $\omegam=0, \omegal=0$ model over the
$\omegam=0.28, \omegal=0.72$ model with the same odds that now favor the 
best-fitting flat-$\Lambda$ model. We emphasize however that we are not suggesting 
that there is evidence for such a shift in magnitude.

For comparison, using the same prior, the odds favoring the best-fit 
flat-$\Lambda$ model (with $\omegam=0.24$ and $\omegal=0.76$) over the 
best-fit open model (with $\omegam=0$ and $\omegal=0$) are $3:1$ after 
examining the R98 data.

The implication of these analyses is clear: the SNe Ia data sets are now
large enough to achieve powerful statistical results. Confidence in
such results will obtain from more detailed investigation of possible
systematic effects.

\section{Constraints on a Time-Variable Cosmological ``Constant"}
\label{quint}

While the restricted one-dimensional models (flat-constant-$\Lambda$ and open) 
discussed in the previous two sections are consistent with most
recent observations, the flat-constant-$\Lambda$ model seems to be in
conflict with a number of observations, including: (1) analyses of the
rate of gravitational lensing of quasars and radio sources by
foreground galaxies which require a rather large $\omegam \geq 0.38$
at 2 $\sigma$ in this model (see, e.g., Falco, Kochanek, \& Mu\~noz
1998); and (2) analyses of the number of large arcs formed by strong
gravitational lensing by clusters (Bartelmann et al. 1998, also see
Meneghetti et al. 2000; Flores, Maller, \& Primack 2000)\footnote{
Note that the constraints on the flat-constant-$\Lambda$ model from
gravitational lensing of quasars (not radio sources) might be less restrictive
than previously thought (see, e.g., Chiba \& Yoshii 1999; Cheng \& Krauss
2000), and semi-analytical analyses of large-arc statistics lead
to a different conclusion (Cooray 1999; Kaufmann \& Straumann 2000).}.

In the near future, measurements of the cosmic microwave background (CMB) 
anisotropy, thought to be generated by zero-point quantum fluctuations 
during inflation (see, e.g., Fischler, Ratra, \& Susskind 1985), will 
provide a tight determination of cosmological parameters. See, e.g., 
Kamionkowski \& Kosowsky (1999), Rocha (1999), Page (1999) and Gawiser \&
Silk (2000) for recent reviews of the field.

While it has been suggested, largely from $\chi^2$ comparisons of CMB 
anisotropy measurements and model predictions (Ganga, Ratra, \& Sugiyama
1996), that a spatially-flat model is favored over an open one (see, e.g., 
Lineweaver 1999; Dodelson \& Knox 2000; Peterson et al. 2000; Page 1999;
Melchiorri et al. 2000; Tegmark \& Zaldarriaga 2000; Knox \& Page 2000; 
Le Dour et al. 2000), 
such suggestions must be viewed as tentative (see discussion in Ratra et al.
1999 and references therein); see however Lange et al. (2000). More reliable
constraints follow from models-based maximum likelihood analyses of CMB 
anisotropy data (see, e.g., G\'orski et al. 1995; Ganga et al. 1997, 1998; 
Rocha et al. 1999). But this method has not yet been applied to enough data 
sets to provide robust statistical constraints.

A spatially-flat model with a time-variable $\Lambda$ can probably be
reconciled with some of the observations that conflict with a large constant 
$\Lambda$ (e.g., Peebles \& Ratra 1988; Ratra \& Quillen 1992; Perlmutter, 
Turner, \& White 1999b; Wang et al. 2000; Efstathiou 1999; Podariu \& Ratra 
2000; Waga \& Frieman 2000). We emphasize, however, that most current 
observational indications are tentative and not definitive.

At present, the only consistent model for a time-variable $\Lambda$ is that
which uses a scalar field ($\phi$) with a scalar field potential $V(\phi)$
(Ratra \& Peebles 1988). In this paper we focus on the favored scalar field
model in which the potential $V(\phi) \propto \phi^{-\alpha}$, $\alpha > 0$, 
at low redshift (Peebles \& Ratra 1988; Ratra \& Peebles 1988)\footnote{
Other potentials have been also considered, e.g., an exponential potential 
(see, e.g., Lucchin \& Matarrese 1985; Ratra \& Peebles 1988; Ratra 1989; 
Wetterich 1995; Ferreira \& Joyce 1998), but such models are inconsistent with 
observational data. A potential $\propto \phi^{-\alpha}$ plays a role in some 
high energy particle physics models (see, e.g., Masiero \& Rosati 1999; 
Albrecht \& Skordis 2000; de la Macorra 1999; Brax \& Martin 2000; Choi 1999). 
Discussions of these and related models are given by Steinhardt, Wang, \& 
Zlatev (1999), Chiba (1999), Amendola (1999), de Ritis et al. (2000), Fujii 
(2000), Holden \& Wands (2000), Bartolo \& Pietroni (2000), and Barreiro, 
Copeland, \& Nunes (2000). \"Ozer (1999), Waga \& Miceli (1999), Battye, 
Bucher, \& Spergel (1999), and Bertolami \& Martins (1999) discuss other 
possibilities.}.
A scalar field is mathematically equivalent to a fluid with a time-dependent speed of sound (Ratra 1991). This equivalence may be used to show that a 
scalar field with potential $V(\phi) \propto \phi^{-\alpha}$, $\alpha > 0$, 
acts like a fluid with negative pressure and that the $\phi$ energy density 
behaves like a cosmological constant that decreases with time. We emphasize 
that in the analysis here we do not make use of the time-independent equation 
of state fluid approximation to the scalar field model for a time-variable 
$\Lambda$, as has been done in a number of recent papers (see discussion in 
Podariu \& Ratra 2000; also see Waga \& Frieman 2000). 

The SNe Ia data also place constraints on a time-variable $\Lambda$. Here 
we consider only spatially flat models. For each SN Ia, there is a locus of 
values of $\alpha$ and $\omegam$ that predict the corrected apparent 
magnitude. These curves define regions of different likelihood in the
$\alpha-\omegam$ plane. Figure 6 shows this plane, with greyscale intensity 
proportional to the binomial likelihood (eq. [1]) using 16 R98 high-$z$ SNe Ia. 

We now compute the posterior odds of the time-variable $\Lambda$ model versus 
the time-independent $\Lambda$ model, allowing only spatially flat 
cosmologies. The prior odds are set as in Section \ref{bayes}, thus we 
penalize complicated models by the prior $P\propto (1/2)^{N+1}$ where $N$ is
the number of parameters. The time-variable $\Lambda$ model has two parameters,
$\alpha$ and $\omegam$, while the flat-constant-$\Lambda$ model has but one. 
Thus the prior odds are $2:1$ in favor of the constant $\Lambda$ model before 
examining the SNe Ia data. We focus on the range $0 \leq \alpha \leq 8$ since 
for larger $\alpha$ the time-variable $\Lambda$ model approaches the 
Einstein-de Sitter one (Peebles \& Ratra 1988). For computational simplicity 
we also focus on the range $0.05 \leq \omegam \leq 0.95$.

To compare the time-variable $\Lambda$ model with the flat-constant-$\Lambda$ 
model, we compute average likelihoods for $\alpha>0$ and $\alpha=0$,
adopting a uniform prior for both parameters over the ranges $0 \leq \alpha 
\leq 8$ and $0.05 \leq \omegam \leq 0.95$. For the R98 data, the ratio of 
average likelihoods is $2:1$ in favor of the constant $\Lambda$ model, 
thus the posterior odds are $3.9:1$. Applying the same analysis to the P99 
data, we find that these data favor the constant $\Lambda$ model by $18:1$ 
over the time-variable model.

If we adopt logarithmic priors for both $\alpha$ and $\omegam$ (in this case 
setting a lower bound $\alpha>0.01$ to the time-variable model --- these 
results are insensitive to any $0.01<\alpha_{\rm min}<0.1$), the R98 data favor 
the constant-$\Lambda$ model by odds of $1.7:1$. However, log priors
cause the P99 data to favor the time-variable $\Lambda$ model by $3.6:1$.
The latter occurs because there are several SNe Ia in the P99 data set whose
brightnesses would be matched by quite small (but non-zero) values of 
$\alpha$ and $\omegam$, so giving larger weight to this region in the 
$\alpha-\omegam$ plane strongly increases the average likelihood for the 
time-variable $\Lambda$ model. The strong dependence of the results on the 
prior distribution for the parameters indicates that better data are needed 
to convincingly constrain these parameters.

We also compare the time-variable $\Lambda$ model to the open one with
$0.05<\omegal<1$ and $\omegal=0$. For the R98 data the posterior odds are
$3.2:1$ and $1.9:1$ in favor of the time-variable $\Lambda$ model when 
we adopt uniform and logarithmic priors, respectively, for $\alpha$ and
$\omegam$. These results motivate further consideration of time-variable 
$\Lambda$ models.

\section{Conclusions}
\label{conclude}

Applications of median statistics that we present in this paper
demonstrate that statistical independence and freedom from systematic
errors are by themselves extremely powerful hypotheses. Perhaps to the
surprise of most who survived Freshman Physics laboratory, we find
that median statistics leads to strong constraints on models even
though this method does not make use of the other two of four
assumptions required for standard $\chi^2$ data analysis, those that
require Gaussianity and knowledge of the errors. When applied to some
of the astronomical data we consider, the median statistics results
are dramatic enough to make one question even the first two hypotheses
--- independence and freedom from systematic error. Median statistics
are relatively robust to bad data but when median statistics yield
such strong results this could be a warning that the
assumptions of independence and freedom from
systematics should be carefully examined.

Median statistics analysis of 331 Hubble constant estimates, from
Huchra's (1999) compilation, yields a value of $H_0=67\kmsmpc$. This value
is quite reasonable and in agreement with many recent estimates,
including those obtained from the R98 SNe Ia data that we examine.
Based on nearly all available data, this is arguably the best available 
current summary of our knowledge of the Hubble constant. 
Such a summary statistic is useful when one needs a consensus value
for a cosmological simulation or similar application or, as in the
case of the Hayden Planetarium, simply to present a value that is
representative of current knowledge (the Planetarium chose
$H_0=70\kmsmpc,\, \omegam=1/3,\, \omegal=2/3$, just one significant figure
for each constant). The formal, purely statistical 95\% 
confidence interval that results from median statistics, $65-69\kmsmpc$,
is indeed narrow, which highlights the power of our assumptions. If they 
were truly independent and free of systematics, the extant estimates of 
$H_0$ would clearly be numerous enough.

Systematic effects do, of course, dominate the error budget for the
Hubble Constant.  The vast majority of the published estimates share
possible systematic uncertainty through the LMC distance scale and/or
calibration of the Cepheid period-luminosity relation (as many as 279
of the 331 could be so affected).  We apply median statistics to the
distribution of different methods for measuring the LMC distance
modulus and find a median value $m-M=18.46$ with 95\% confidence
limits $18.26<(m-M)<18.64$. This range of distance moduli implies that
the systematic error in our estimate of the median of $H_0$ could be
as large as 7.5\% in either direction (95\% limits). Grouping the 331
estimates of the Hubble constant by method and applying median
statistics to the distribution of methods, we infer that the 95\%
confidence range of systematic error due to differences between
methods is 7\% (the median is $H_0=70\kmsmpc$ with 95\% range $64.5 -
 74.5 \kmsmpc$). To be conservative, we take the somewhat larger of these
two estimates of systematic uncertainty and quote a total error budget
on the true median of $H_0$ of $H_0=67 \pm 2 (95\% \ {\rm statistical})
\pm 5 (95\% \ {\rm systematic})\kmsmpc$. Thus, systematic errors clearly
dominate over the purely statistical errors.

Of some interest is the dependence, or near lack thereof, of median
statistics of $H_0$ on the authors of the papers or the year of
publication. Completely excluding all the work of any of the
best-known investigators or groups -- Sandage, Tammann, and
collaborators, de Vaucouleurs or van den Bergh,  or the HST Key
Project -- has at most a $2\kmsmpc$ effect on the median. 
The set of estimates attributed to none of these groups has median
$H_0=68\kmsmpc$; this value is also the median of the medians from each group.
Recent $H_0$ estimates (post 1990 or post 1996) differ only slightly from the
median of all estimates, shifting the median to $H_0=65\kmsmpc$ with
confidence limits that include the value estimated from the full data
set.

Our analyses of constraints on $\omegam$ and $\omegal$ from recently
published high-$z$ SNe Ia data from R98 and P99 generally support the
conclusions of these groups.  Although our results differ in detail,
our median statistics prefer the same region in the ($\omegam$,
$\omegal$) plane as did these earlier analyses.  Because we abandon
the assumption of Gaussianity, the statistical power of our results is
somewhat smaller. If the assumption of Gaussianity is valid, then
somewhat stronger constraints (with confidence similar to limits found by R98
and P99) could be obtained but these would not be identical to those
of R98 and P99 because we assume a different prior.

In agreement with R98 and P99, the $\omegam = 1$ Einstein-de Sitter
model is strongly ruled out. The reason for this strong result is
simply that the majority of the SNe Ia are too faint for the model.
Using only the binomial likelihoods that the observed SNe Ia are too
bright/faint for a given model, we find that the 16 R98 high-$z$ SNe
Ia rule out the Einstein-de Sitter model at the 99.6\% confidence
level.  A similar analysis rules out $\omegal=0$ models at 89\%.

We apply a more complete Bayesian treatment to the 16 R98 SNe Ia, including 
appropriate priors for $\omegam$ and $\omegal$, and for models with varying 
numbers of free parameters. The posterior probability that $\Lambda > 0$ is 
between 70 and 89\%, depending on how we bound the parameter space using prior 
information (compare Tables 4 and 6). The posterior probability of an open 
universe is about 47\% and the probability of a flat universe is either 51 or 
38\%. These results differ in detail from those of R98 (and a similar 
conclusion holds for the results of P99), whose analysis used a uniform prior
for $\omegam$ and made no allowance for the zero- or one-parameter models. 
The constraints on $\omegal$ are not sensitive to our use of a logarithmic 
prior for $\omegam$, although the uniform prior does strongly discriminate 
against low $\omegam$ models and significantly increases the odds of a flat 
model over an open one (also see Podariu and Ratra 2000).

To determine the significance of constraints  on $\omegam$ and $\omegal$ 
from a larger data set, we apply median statistics to the 42 high-$z$ SNe Ia
reported by P99. Here we simply count the number of SNe Ia that lie 
brighter/fainter than predicted by different models and compute the binomial 
likelihoods of these events. The likelihood of the best fitting flat-$\Lambda$ 
model (with $\omegam=0.28$ and $\omegal=0.72$) is 366 times that of the 
best-fitting open model (with $\omegam=0$ and $\omegal=0$). Thus, if a priori 
we regarded the flat-$\Lambda$ model and the open model as equally likely 
($1:1$), then after examining the P99 data we should favor the flat-$\Lambda$
model by odds of $366:1$. (A similar analysis of the R98 data results in odds
of $3:1$ in favor of the flat-$\Lambda$ model.) That we can achieve such 
dramatic constraints from median statistics alone indicates that it might be
wise to carefully examine the possible effects of systematic errors. 
Although we do not mean to suggest that there is evidence for such an effect,
we caution that a systematic shift of only $0.14$ mag would reverse these odds.

Using similar techniques, we use the SNe Ia to evaluate the posterior
probabilities of a time-variable cosmological ``constant'' compared to a 
flat-constant-$\Lambda$ model. Using uniform priors for the distribution 
of the parameters $\alpha$ and $\omegam$, the R98 and P99 data favor the
constant $\Lambda$ model over the time-variable $\Lambda$ one by posterior 
odds of $3.9:1$ and $18:1$, respectively. If we adopt logarithmic 
priors for the parameters, the R98 data favor the constant $\Lambda$ model 
by somewhat smaller odds, $1.7:1$, but the P99 data actually favor a time-variable $\Lambda$ by $3.6:1$. Similar analysis shows that the R98 data 
mildly favors a time-variable $\Lambda$ model over an open universe with 
$\Lambda=0$, by posterior odds $3.2:1$ or $1.9:1$ assuming uniform or 
logarithmic priors, respectively, on $\alpha$ and $\omegam$. We conclude that
the data in hand are not good enough to convincingly constrain these parameters.

Given the simplicity of median statistics and their freedom from the
sometimes-questionable assumption of Gaussianity, we find it surprising
that such methods have not been applied more frequently. At the very
least, this approach is useful for early analyses of data sets, before
one has gathered the evidence to justify methods that require stronger
hypotheses. As our examples illustrate, when applied even to larger
data sets, median statistics provide a check on more complicated methods.
When the results of median statistics seem questionable, analyses that
rely on a larger number of assumptions are likely to be even more in
doubt.  We suggest that one follow the advice of Zeldovich. Take the median!

\bigskip

We acknowledge valuable discussions with I. Wasserman. We thank J.
Huchra for his compilation of Hubble constant estimates and R.
Marcialis for providing the Pluto-Charon data. We are indebted to the
referee, B. Gibson, for detailed comments which helped improved the
paper. We also thank C. Dudley, A. Gould, A. Riess and L. Weaver for
helpful comments on the manuscript.  JRG acknowledges support from NSF
grant AST-9900772. MSV acknowledges support from the AAS Small
Research Grant program, from John Templeton Foundation grant
938-COS302, and from NSF grant AST-0071201. SP and BR acknowledge
support from NSF CAREER grant AST-9875031.

\clearpage

\begin{deluxetable}{lrr}
\tablecaption{Hubble Constant Medians by Type}
\tablewidth{0pt}
\tablehead{
\colhead{Type of Estimate} &
\colhead{Number} &
\colhead{Median}
}
\startdata
No Type & 1 & 85\\
SNe II & 2 & 66.5\\
Global Summary & 70& 70\\
B Tully-Fisher & 21 & 57\\
CMB fit & 1 & 30\\
$D_n-\sigma$ & 9 & 75\\
SB Fluctuations & 8 & 82\\
Glob. Cluster LF & 12 & 76.5\\
IR Tully-Fisher & 16 & 85\\
Irvine meeting & 5 & 67 \\
Grav. Lensing & 26 & 64.5\\
Novae & 3 & 69\\
Other & 54 & 70\\
Plan. Nebulae LF & 3 & 87\\
I, R Tully-Fisher & 11 & 74\\
SNe I& 55 & 60\\
Tully-Fisher & 9 & 73\\
Sunyaev-Zeldovich & 25 & 55
\enddata
\end{deluxetable}

\begin{deluxetable}{llr}
\tablecaption{Binomial Probabilities of $\omegam$ (Assuming $\omegal=0$) from 
16 High-$z$ SNe Ia}
\tablewidth{0pt}
\tablehead{
\colhead{$\omegam$ Estimated from SN} &
\colhead{Probability of TM [\%]} &
\colhead{Relative Likelihood}
}
\startdata
 &  0.00153 & 1 \\[-4pt]
5.96 &  &  \\[-4pt]
 & 0.0244 & 16 \\[-4pt]
1.68 & & \\[-4pt]
 & 0.183 & 120 \\[-4pt]
0.656 & & \\[-4pt]
 & 0.854 & 560 \\[-4pt]
0.206 & & \\[-4pt]
 & 2.78 & 1,820 \\[-4pt]
0.0426 & & \\[-4pt]
 & 6.67 & 4,368 \\[-4pt]
$-$0.0136 & & \\[-4pt]
 & 12.2 & 8,008 \\[-4pt]
$-$0.165 & & \\[-4pt]
 & 17.5 &  11,440 \\[-4pt]
$-$0.266 & & \\[-4pt]
 & 19.6 & 12,870 \\[-4pt]
$-$0.303 & & \\[-4pt]
 & 17.5 & 11,440 \\[-4pt]
$-$0.310 & & \\[-4pt]
 & 12.2 & 8,008 \\[-4pt]
$-$0.349 & & \\[-4pt]
 & 6.67 & 4,368 \\[-4pt]
$-$0.724 & & \\[-4pt]
 & 2.78 & 1,820 \\[-4pt]
$-$1.33 & & \\[-4pt]
 & 0.854 & 560 \\[-4pt]
$-$1.60 & & \\[-4pt]
 & 0.183 & 120 \\[-4pt]
$-$1.62 & & \\[-4pt]
 & 0.0244 & 16 \\[-4pt]
$-$2.53 & & \\[-4pt]
 & 0.00153 & 1
\enddata
\end{deluxetable}

\begin{deluxetable}{ccc}
\tablecaption{Hypothetical Early 1960's Bayesian Analysis}
\tablewidth{0pt}
\tablehead{
\colhead{Model} &
\colhead{Prior Probability [\%]} &
\colhead{Posterior Probability [\%] after 16 SNe}
}
\startdata
Steady-state & 33.3 & 67.3\\
\sidehead{$\omegal=0$ with:}
$1 < \omegam < 4$ & 10.5 & 0.642 \\
$\omegam = 1$ & 33.3 & 4.44 \\
$0.2 < \omegam < 1$ & 12.2 &  6.35 \\
$0.05 < \omegam < 0.2$ & 10.5 &  21.3
\enddata
\end{deluxetable}

\begin{deluxetable}{ccc}
\tablecaption{2000 Bayesian Analysis with $P(\omegam)\propto 
d\omegam /\omegam$}
\tablewidth{0pt}
\tablehead{
\colhead{Model} &
\colhead{Prior Probability [\%]} &
\colhead{Posterior Probability [\%] after 16 SNe}
}
\startdata
\sidehead{$\omegal = 0$ with:}
$1 < \omegam < 4$ & 4.52 & 0.34 \\
$\omegam = 1$ &  57.1 & 9.37 \\
$0.2 < \omegam < 1$ & 5.25 & 3.92 \\
$0.05 < \omegam < 0.2$ & 4.52 & 13.33 \\
\sidehead{Flat-$\Lambda$ ($\omegam + \omegal = 1$) with:}
$0 < \omegal < 0.95$ & 6.97 & 41.53 \\
$-1 < \omegal < 0$ &  7.33 & 0.60 \\
\sidehead{Two-parameter ($0.05 < \omegam < 4$, $-1 < \omegal < 1$):}
Open \& $\omegal > 0$ & 3.34 & 27.48 \\
Closed \& $\omegal > 0$ & 3.81 & 1.15 \\
Open \& $\omegal < 0$ & 5.51 & 2.23 \\
Closed \& $\omegal < 0$ & 1.63 & 0.05 \\
\enddata
\end{deluxetable}

\begin{deluxetable}{ccc}
\tablecaption{Summary of 2000 Bayesian Analysis with $P(\omegam)\propto 
d\omegam /\omegam$}
\tablewidth{0pt}
\tablehead{
\colhead{Model/Parameter Range} &
\colhead{Posterior Probability [\%] after 16 SNe}
}
\startdata
$\Lambda > 0$ & 70.16 \\
$\Lambda = 0$ & 26.96 \\
$\Lambda < 0$ & 2.88 \\
 & & \\
Flat & 51.50 \\
Open & 46.96 \\
Closed & 1.54 \\
\enddata
\end{deluxetable}

\begin{deluxetable}{ccc}
\tablecaption{Restricted 2000 Bayesian Analysis with $P(\omegam)\propto 
d\omegam /\omegam$}
\tablewidth{0pt}
\tablehead{
\colhead{Model} &
\colhead{Prior Probability [\%]} &
\colhead{Posterior Probability [\%] after 16 SNe}
}
\startdata
\sidehead{$\omegal = 0$ with:}
$0.2 < \omegam < 1$ & 17.9 & 2.58 \\
$0.05 < \omegam < 0.2$ & 15.4 & 8.75 \\
\sidehead{Flat-$\Lambda$ ($\omegam + \omegal = 1$) with:}
$0 < \omegal < 0.95$ & 33.3 & 38.35 \\
\sidehead{Two-parameter ($0.05 < \omegam < 1$, $0 < \omegal < 1$):}
Open \& $\omegal > 0$ & 22.8 & 36.16 \\
Closed \& $\omegal > 0$ & 10.5 & 14.16
\enddata
\end{deluxetable}

\begin{deluxetable}{ccc}
\tablecaption{2000 Bayesian Analysis with $P(\omegam)\propto d\omegam$}
\tablewidth{0pt}
\tablehead{
\colhead{Model} &
\colhead{Prior Probability [\%]} &
\colhead{Posterior Probability [\%] after 16 SNe}
}
\startdata
\sidehead{$\omegal = 0$ with:}
$1 < \omegam < 4$ & 10.96 & 0.94 \\
$\omegam = 1$ &  57.1 & 15.10 \\
$0.2 < \omegam < 1$ & 2.86 & 2.63 \\
$0.05 < \omegam < 0.2$ & 0.57 & 2.47 \\
\sidehead{Flat-$\Lambda$ ($\omegam + \omegal = 1$) with:}
$0 < \omegal < 0.95$ & 6.97 & 59.20 \\
$-1 < \omegal < 0$ &  7.31 & 0.82 \\
\sidehead{Two-parameter ($0.05 < \omegam < 4$, $-1 < \omegal < 1$):}
Open \& $\omegal > 0$ & 0.81 & 7.73 \\
Closed \& $\omegal > 0$ & 6.33 & 10.00 \\
Open \& $\omegal < 0$ & 2.63 & 0.87 \\
Closed \& $\omegal < 0$ & 4.53 & 0.20 \\
\enddata
\end{deluxetable}

\begin{deluxetable}{ccccc}
\tablecaption{Binomial Likelihoods using 42 High-$z$ SNe Ia}
\tablewidth{0pt}
\tablehead{
\colhead{$\omegam$} &
\colhead{$\omegal$} &
\colhead{Too Bright} &
\colhead{Too Faint} &
\colhead{Relative Likelihood}
}
\startdata
1     & 0    & 4  & 32 & 0.000076 \\
0     & 0    & 10 & 32 & 1 \\
0.28  & 0.72 & 21 & 21 & 366 \\
0     & 1    & 31 & 11 & 2.9
\enddata
\end{deluxetable}

\clearpage

\centerline{\bf FIGURE CAPTIONS}

% Fig 1
\figcaption[]{Relative likelihood of the true median of $H_0$ from 331
published estimates compiled by Huchra (1999). Estimates were
tabulated to the nearest integer and include many that are
identical. The bins of likelihood are centered on these integral
values. Each bin includes the sum of binomial likelihoods that are
sandwiched between the identical estimates, one-half the likelihood
that $H_0$ lies between this and the next smaller unique estimate,
and one-half the likelihood of $H_0$ between this and the next
larger unique estimate. The median of the estimates is $67\kmsmpc$
with 95\% (2 $\sigma$) statistical confidence interval of
$65-69\kmsmpc$, computed using the integral over the tails of the
binomial probability distribution. Systematic 
uncertainty is approximately $\pm 5\kmsmpc \ (95\% \ {\rm limits})$.}

% Fig 2
\figcaption[]{Relative likelihood of the true median of estimates of the mass 
of the Pluto-Charon system (data from Marcialis 1997) from all measurements 
(unhatched area) and 
from those published after 1950 (hatched area). The large number of early 
erroneous estimates pulls the median far from the modernly-accepted value. 
Even median statistics can be swayed by too much bad data. Restricting the 
analysis to post-1950 estimates yields a median that is identical to the 
modernly-accepted value; the median is 0.00246 Earth masses (M$_{\rm E}$), 
with 95\% confidence range from 0.00236 to 0.08 M$_{\rm E}$.}

% Fig 3
\figcaption[]{($\omegam$, $\omegal$) loci for 16 R98 high-$z$ SNe Ia. The 
curve for each SN is the set of values of $\omegam$ and $\omegal$ that 
predict that SN's corrected apparent magnitude. To compute the likelihood 
of a particular model, count the number of curves that lie above and below 
that model's location in this plane and apply the binomial theorem. Figure 
3$a$ shows loci over a large range of $\log\omegam, \omegal$. Models with 
parameter values in the upper left hand corner region bounded by the 
short-dashed curve do not have a big bang. The horizontal dot-dashed line 
demarcates models with a zero $\Lambda$ and the long-dashed curve 
indicates spatially-flat models. Figure 3$b$ shows only a selected range of 
$\omegam, \omegal$, with linear axes. The long-dashed diagonal line indicates 
spatially-flat models. In this linear plot, 2 SNe Ia lie below the bottom of 
the figure and 4 lie above the top.}

% Fig 4
\figcaption[]{Relative likelihood of $\omegam$, $\omegal$ from the 16
R98 high-$z$ SNe Ia. Beginning with the set of supernova loci shown in
Figure 3, the greyscale intensity at each point in the ($\omegam$,
$\omegal$) plane is set proportional to the binomial likelihood of the
observed number of supernovae being brighter/fainter than expected for
that model. Conventions, including definitions of dot-, long-, and 
short-dashed curves, are as described in the caption of Figure 3. The
solid lines are 1, 2, and 3 $\sigma$ likelihood contours computed using a 
$\chi^2$ analysis similar to that used by R98 (see Podariu \& Ratra 2000).
In Figure 4$a$ the likelihood is set to zero in the ``No Big Bang'' region.}

% Fig 5
\figcaption[]{Relative likelihood of $\omegam$ from the 16 R98 high-$z$
SNe Ia for spatially flat models with $\omegam + \omegal = 1$. These are 
the relative likelihoods of points along the ``flat universe'' long-dashed 
curve in Figure 4$a$.}

% Fig 6
\figcaption[]{Relative likelihood of the parameters $\alpha$ and $\omegam$
for time-variable $\Lambda$ models, using the 16 R98 high-$z$ SNe Ia to
estimate the likelihoods. Greyscale intensities are computed as described in
the captions of Figures 3 and 4. The solid lines are 1, 2, and 3 $\sigma$ 
confidence contours computed using a $\chi^2$ analysis (Podariu \& Ratra 2000);
the first curve in the lower left-hand corner is a 2 $\sigma$ contour and
one of the 1 $\sigma$ contours is obscured by the shading.}

\clearpage

\begin{figure}
\resizebox{\textwidth}{!}{\includegraphics{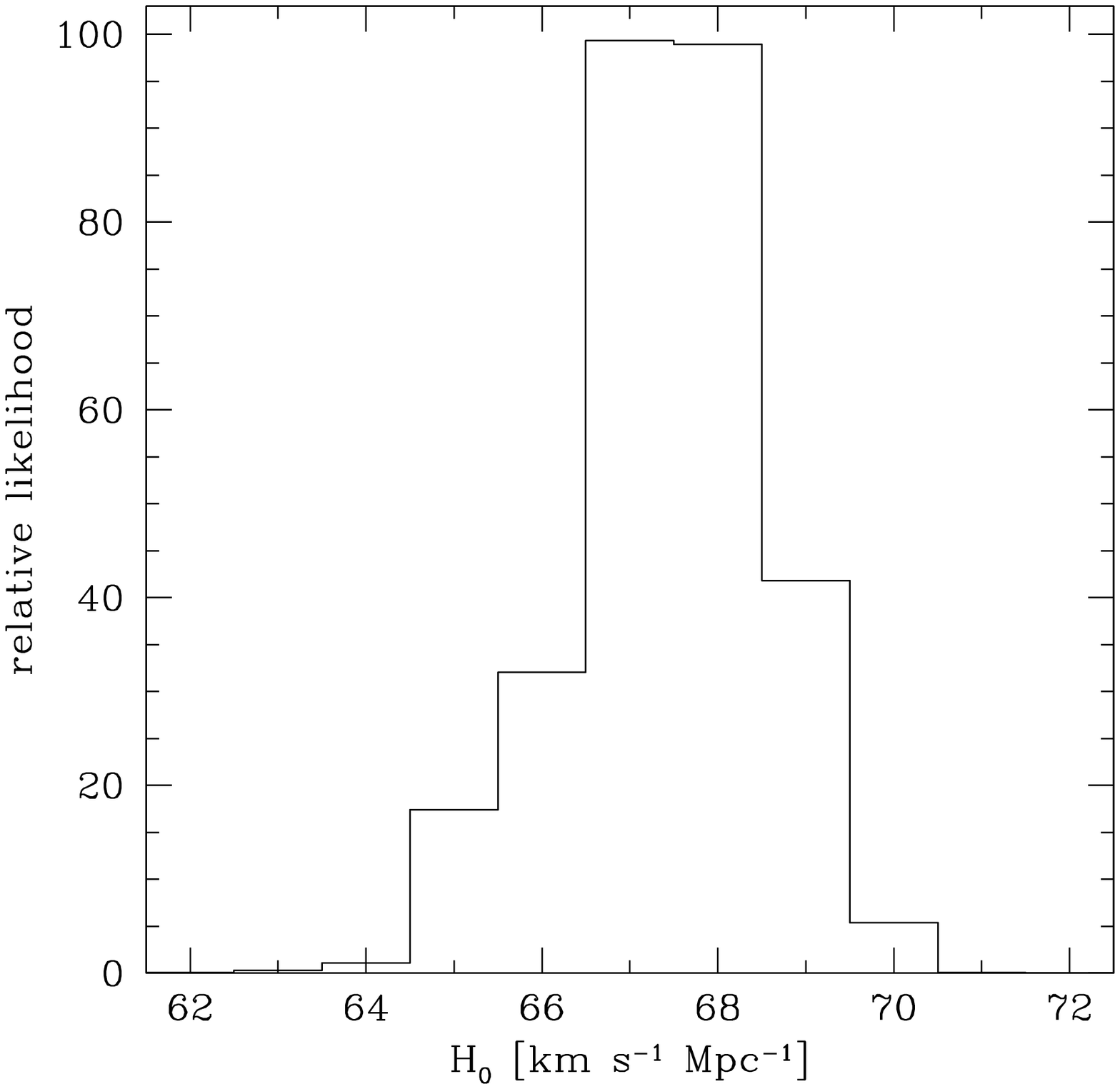}}
Figure 1
\end{figure}
\clearpage

\begin{figure}
\resizebox{\textwidth}{!}{\includegraphics{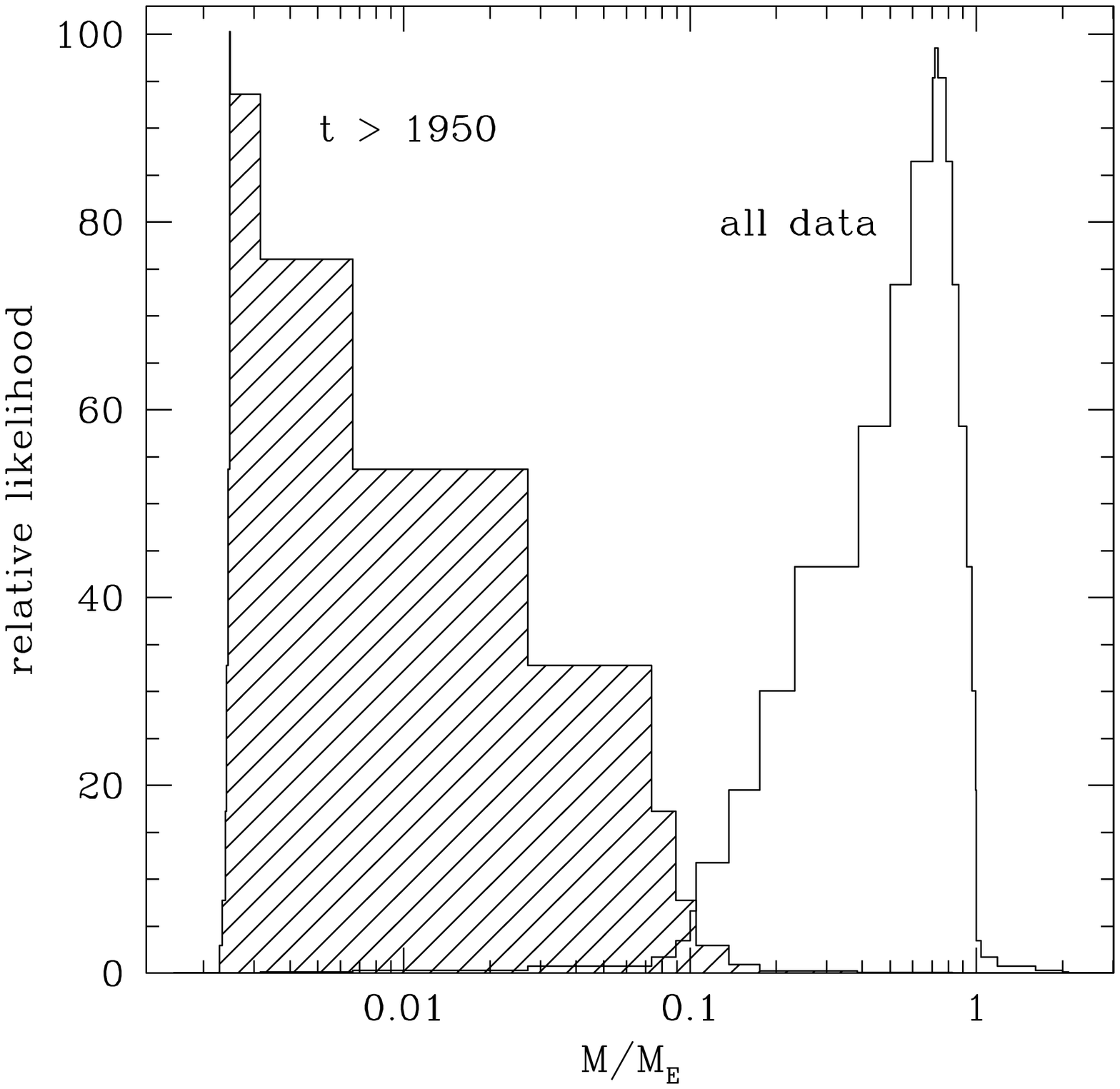}}
Figure 2
\end{figure}
\clearpage

\begin{figure}
\resizebox{\textwidth}{!}{\includegraphics{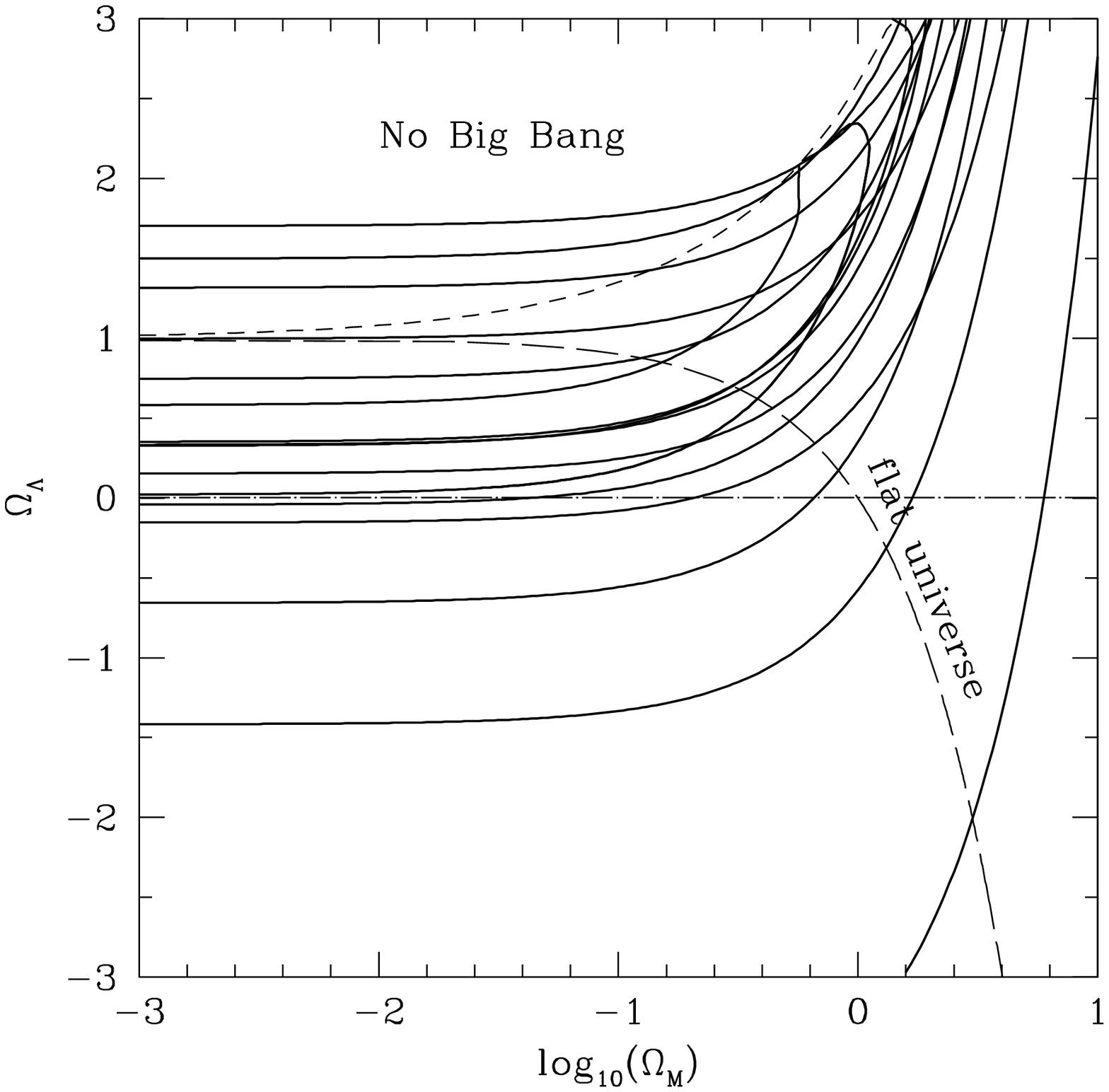}}
Figure 3a
\end{figure}
\clearpage

\begin{figure}
\resizebox{\textwidth}{!}{\includegraphics{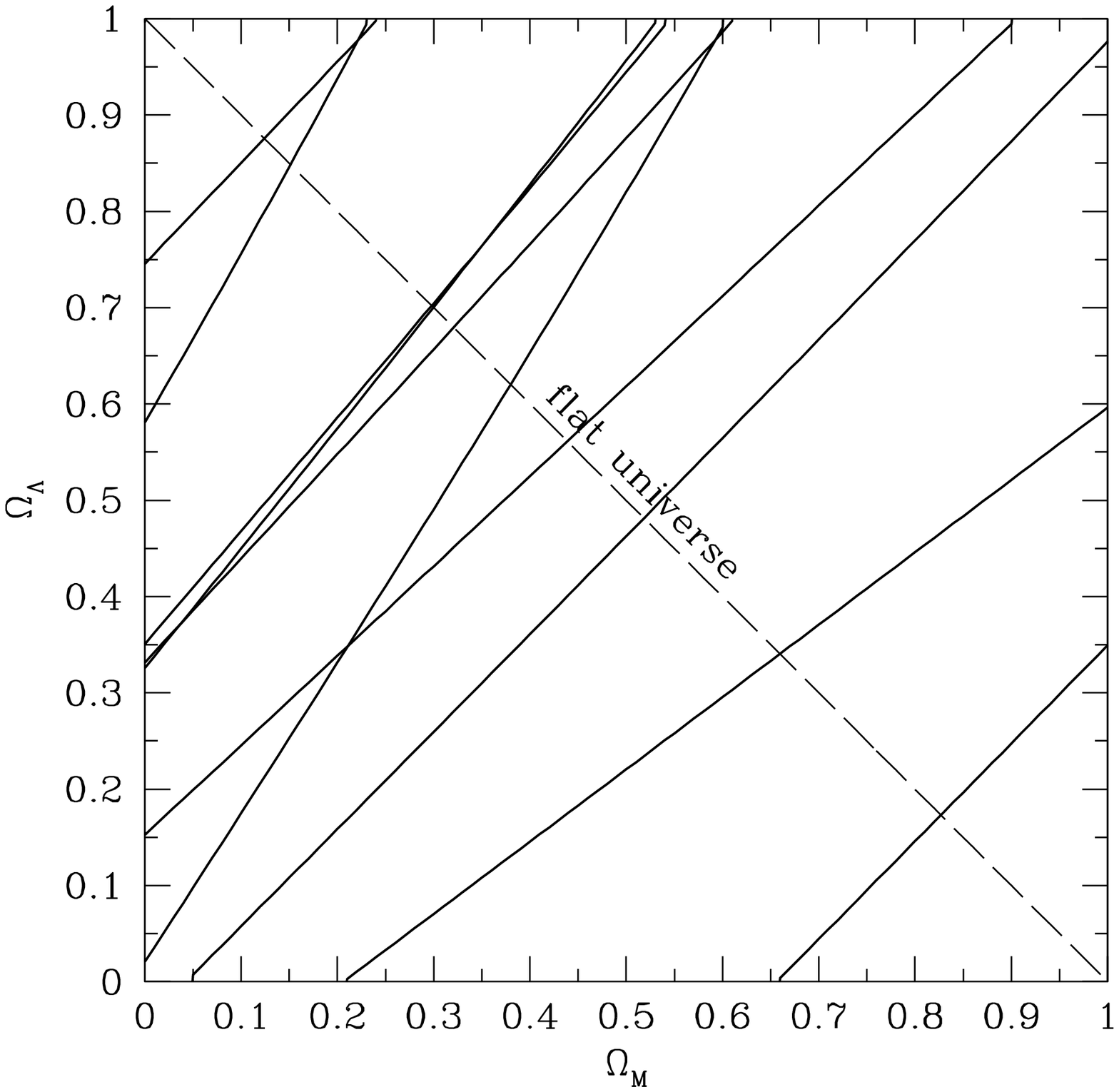}}
Figure 3b
\end{figure}
\clearpage

\begin{figure}
\resizebox{\textwidth}{!}{\includegraphics{figure4a.ps}}
Figure 4a
\end{figure}
\clearpage

\begin{figure}
\resizebox{\textwidth}{!}{\includegraphics{figure4b.ps}}
Figure 4b
\end{figure}
\clearpage

\begin{figure}
\resizebox{\textwidth}{!}{\includegraphics{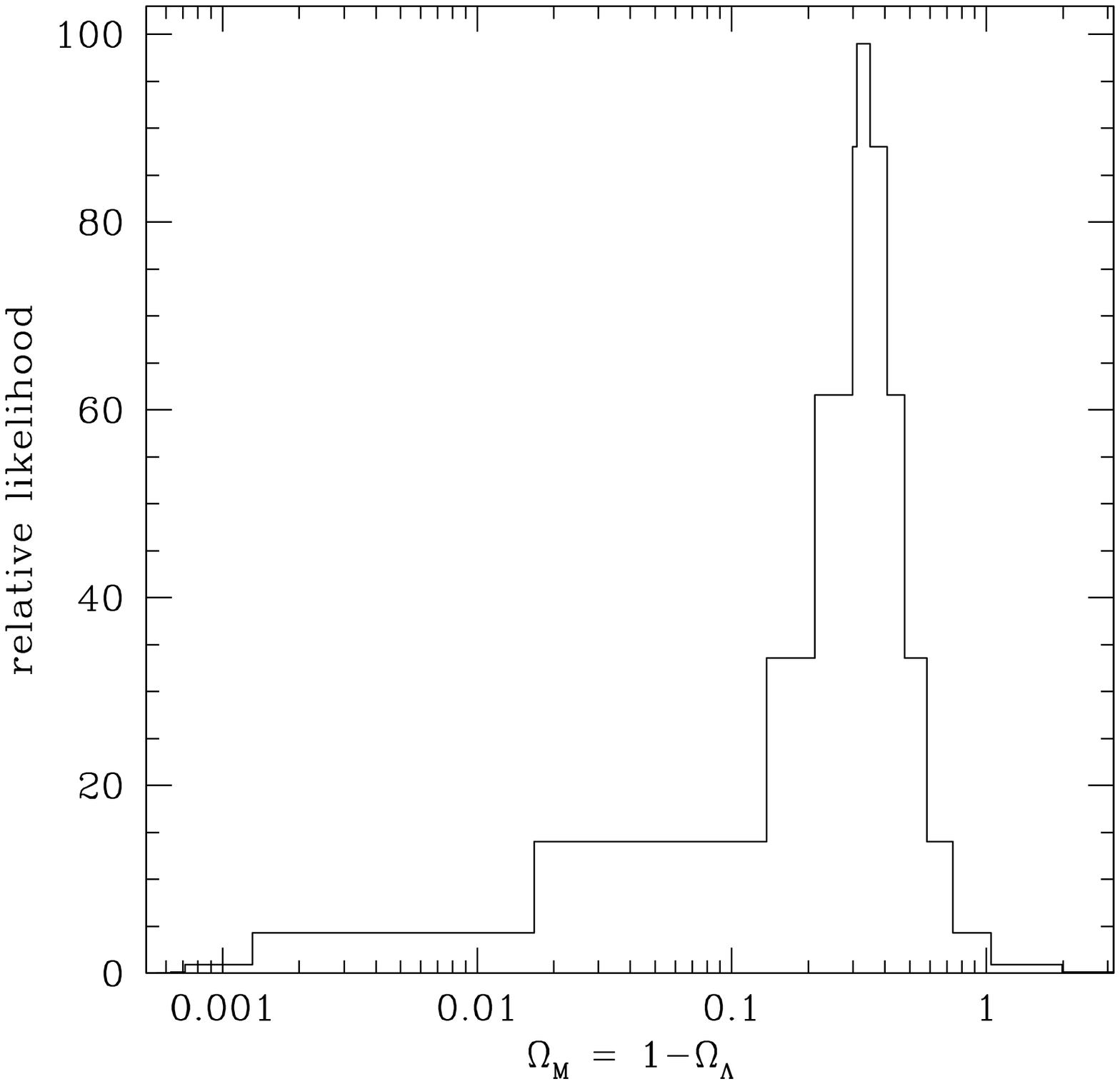}}
Figure 5
\end{figure}
\clearpage

\begin{figure}
\resizebox{\textwidth}{!}{\includegraphics{figure6.ps}}
Figure 6
\end{figure}
\clearpage

\end{document}